\documentclass{aa}
\usepackage[dvips]{graphics}
\newcommand{\tg}{T_{\rm g}}
\newcommand{\tr}{T_{\rm r}}

\newcommand{\hhdp}{H$_2$D$^+$}

\newcommand{\beq}{\begin{equation}}
\newcommand{\enq}{\end{equation}}

\begin{document}

\thesaurus{02(12.05.1; 02.01.4; 02.13.5)}
\title{The chemistry of the early Universe}
\author{Daniele Galli and Francesco Palla}
\institute{
Osservatorio Astrofisico di Arcetri, 
Largo E. Fermi 5, 
I-50125 Firenze, 
Italy}
\offprints{D. Galli}
\date{Received date; accepted date}
\maketitle
\begin{abstract}

The process of molecule formation in the primordial gas is considered
in the framework of Friedmann cosmological models from redshift
$z=10^4$ to $z=0$. First, a comprehensive analysis of 87 gas phase
reaction rates (both collisional and radiative) relevant in the
physical environment of the expanding universe is presented and
critically discussed.  On this basis, calculations are carried out of
the abundance of 21 molecular species as function of redshift, for
different values of the cosmological parameters $\Omega_0$, $\eta$ and
$H_0$, evaluating consistently the molecular heating and cooling due to
H$_2$, HD and LiH molecules.  One of the major improvements of this
work is the use of a better treatment of H recombination that leads to
a reduction of a factor 2--3 in the abundance of electrons and H$^+$ at
freeze-out, with respect to previous studies. The lower residual
ionization has a negative effect on the chemistry of the primordial gas
in which electrons and protons act as catalysts in the formation of the
first molecules.  We find that in the standard model ($h=0.67$,
$\eta_{10}=4.5$, $\Omega_0=1$ and [D/H] $=4.3\times 10^{-5}$), the
residual fractional ionization at $z=1$ is $[{\rm e/H}]=3.02\times
10^{-4}$, and the main molecular species fractional abundances
$[{\rm H}_2/{\rm H}]=1.1\times 10^{-6}$, $[{\rm HD/H}_2]=1.1\times
10^{-3}$, $[{\rm HeH}^+/{\rm H}]=6.2\times 10^{-13}$, $[{\rm
LiH}^+/{\rm H}]=9.4\times 10^{-18}$ and $[{\rm LiH/LiH}^+]=7.6\times
10^{-3}$.  We devise a reduced chemical network that reproduces with
excellent accuracy the numerical results of the complete model and
allows to  follow the chemical compositions and the thermal properties
of a primordial gas in the presence of an external radiation field.
Finally, we provide accurate cooling functions of H$_2$, HD and LiH in
a wide range of density and temperature that can be conveniently used
in a variety of cosmological applications.

 
\end{abstract}
 
\section{Introduction}

The study of molecule formation in the post-recombination epoch has
grown considerably in recent years. Saslaw \& Zipoy (1967) and Peebles
\& Dicke (1968) were the first to realize the importance of gas phase
reactions for the formation of the simplest molecule, H$_2$. They
showed that trace amounts of molecular hydrogen, of order
10$^{-6}$--10$^{-5}$, could indeed form via the intermediaries species
H$_2^+$ and H$^-$ once the radiation field no longer contained a high
density of photons with energies above the threshold of dissociation
(2.64 and 0.75 eV, respectively).

The presence of even a trace abundance of H$_2$ is of direct relevance
for the cooling properties of the primordial gas which, in its absence,
would be an extremely poor radiator: cooling by Ly-$\alpha$ photons is
in fact ineffective at temperatures $\la 8000$~K, well above the matter
and radiation temperature in the post-recombination era.  Since the
evolution of primordial density fluctuations is controlled by the
ability of the gas to cool down to low temperatures, it is very
important to obtain a firm picture of the chemistry of the dust-free
gas mixture, not limited to the formation of H$_2$, but also to other
molecules of potential interest. In this regard, Lepp \& Shull (1984) 
and Puy et al. (1993) have
computed the abundances of H$_2$, HD and LiH as a function of redshift
for various cosmological models. Although the final abundances of H$_2$
and HD agree in the two calculations, their evolution with redshift is
markedly different, since the epoch of formation varies by a factor of
$\sim 2$. Also, the LiH abundance shows a large discrepancy of about
two orders of magnitude.

More recently, Palla et al. (1995) have analyzed the effects on the
chemistry of the pregalactic gas of a high primordial D abundance in
the light of the controversial results obtained towards high redshift
quasars (see e.g. Tytler \& Burles 1997).  They found that the
abundance of H$_2$ is rather insensitive to variations in the
cosmological parameters implied by a factor of $\sim 10$ enhancement of
primordial [D/H], while HD and LiH abundances vary by larger amounts.
However, the abundance of LiH, obtained with simple estimates of the
radiative association rate, was largely overestimated. Because of the
potential relevance of the interaction of LiH molecules with the cosmic
background radiation (CBR) (Maoli et al. 1994), a proper treatment of
the lithium chemistry was necessary.  Dalgarno et al. (1996) and
Gianturco \& Gori Giorgi (1996a,b) provided accurate
quantum-mechanical calculations of the main reaction rates.  The
chemistry of lithium in the early universe has been then studied by
Stancil et al. (1996) and Bougleux \& Galli (1997).
Finally, useful reviews of the chemistry of the early universe can be
found in Dalgarno \& Lepp (1987), Black (1991), Shapiro (1992), and
Abel et al. (1997).  The latter two studies, in particular, focus on
the nonequilibrium H$_2$ chemistry in radiative shocks which is thought
to be of primary importance during the gravitational collapse of
density fluctuations (see also Anninos et al. 1997).

In spite of such a wealth of specific studies, a comprehensive analysis
of the subject and a critical discussion of the reaction paths and
rates are still lacking. To overcome this limitation, in this paper we
present a complete treatment of the evolution of {\em all} the
molecular and atomic species formed in the uniform pregalactic medium
at high redshifts ($z<10^4$).  The structure of the paper is as
follows: in Sect.~2 we describe the H, D, He, and Li chemistry, with
a critical discussion of the most important rates; the evolutionary
models are presented in Sect.~3, and the results for the standard
model and the dependence on the cosmological parameters are given in
Sect.~4; Sect.~5 introduces a minimal model which highlights the
dominant reactions for the formation of H$_2$, HD, HeH$^+$, LiH and
LiH$^+$; a comparison with the results of previous studies is given in
Sect.~6, and the conclusions are summarized in Sect.~7. Also, the
Appendix provides the collisional excitation coefficients for HD and
H$_2$ and cooling function of H$_2$, HD and LiH which are needed for 
the computation of the thermal evolution of the primordial gas.

\section{The chemical network}

The chemical composition of the primordial gas consists of e$^-$, H,
H$^+$, H$^-$, D, D$^+$, He, He$^+$, He$^{++}$, Li, Li$^+$, Li$^-$,
H$_2$, H$_2^+$, HD, HD$^+$, HeH$^+$, LiH, LiH$^+$, H$_3^+$, and
H$_2$D$^+$. The fractional abundaces of these species are calculated as
function of redshift, starting at $z=10^4$ where He, H, D, and Li are
fully ionized. The network includes 87 reactions with the rate
coefficients taken from the most recent theoretical and experimental
determinations.  Reactions rates are listed separately for each atomic
element in Tables~1 to 4. In each Table, column 2 gives the reaction,
and column 3 the rate coefficient (in cm$^3$~s$^{-1}$ for collisional
processes, in s$^{-1}$ for photo-processes). The gas and radiation
temperatures are indicated by $\tg$ and $\tr$, respectively. The
temperature range of validity of the rate coefficients and remarks on
the rate are given in column 4, together with some remarks on how the
rate was obtained. When reaction rates were not available, we have
integrated the cross sections and fitted the results with simple
analytical expressions (generally accurate to within 10\%).  The last
column gives the reference for the rate or the cross section adopted.

The assessment of the accuracy of each rate while in principle
extremely valuable, is hampered by the large number of reactions
considered. In this work, we prefer to limit the discussion to those
rates that are critical for a correct estimate of the final molecular
abundances, without omitting any important process.  In doing so, we
will consider the reactions that enter in what will be called below the
{\it minimal model}, i.e.  the reduced set of processes that are needed
to compute the evolution of the main molecular species H$_2$, H$_2^+$,
HD, HeH$^+$, LiH and LiH$^+$. In particular, for hydrogen we compare
our rates to those used by Abel et al. (1997) in their extensive
compilation.  For D, He and Li we will refer to rates found in the
literature as well as in databases.  For ease of comparison, we show
the data from experiments/theory, our numerical fit, and the rates
adopted by other authors in Fig.~1 for H, in Fig.~2 for D and He, and
in Fig.~3 for Li. In this way, one can immediately appreciate the
degree of uncertainty associated with each rate.

\subsection{Hydrogen chemistry}

\begin{table*}[t]
\caption{\sc Reaction rates for Hydrogen species}
\vspace{1em}
\begin{flushleft}
\begin{tabular}{lllll}
\hline
   &  reaction   &   rate (cm$^3$~s$^{-1}$ or s$^{-1}$)   & notes &  reference \\
\hline
   &             &                            &       &            \\
H1) & {\bf H$^+$ + e $\rightarrow$ H + $\gamma$} & $R_{c2}$  & see text & \\
H2) & {\bf H + $\gamma$ $\rightarrow$ H$^+$+ e} &  $R_{2c}$ & see text & \\
H3) & {\bf H + e $\rightarrow$ H$^-$ + $\gamma$}
   & $1.4\times 10^{-18}\tg^{0.928}\exp\left(-\frac{\tg}{16200}\right)$
   & fit
   & DJ \\
H4)& {\bf H$^-$ + $\gamma$ $\rightarrow$ H + e}
   & $1.1\times 10^{-1}\tr^{2.13}\exp{\left(-\frac{8823}{\tr}\right)}$
   & fit 
   & DJ  \\
H5) & {\bf H$^-$ + H $\rightarrow$ H$_2$ + e}
   & $1.5\times 10^{-9}$
   & $\tg\leq 300$
   &  \\
   &
   & $4.0\times 10^{-9}\tg^{-0.17}$
   & $\tg > 300$, fit
   & LDZ \\
H6) & H$^-$ + H$^+$ $\rightarrow$ H$_2^+$ + e
   & $6.9\times 10^{-9}\tg^{-0.35}$
   & $\tg\leq 8000$
   &  \\
   &
   & $9.6\times 10^{-7}\tg^{-0.9}$
   & $\tg > 8000$, fit
   & Po \\
H7)& {\bf H$^-$ + H$^+$ $\rightarrow$ 2H }
   & $5.7\times 10^{-6}\tg^{-0.5}+6.3\times 10^{-8}-$
   &
   &  \\
   &
   & $9.2\times 10^{-11}\tg^{0.5}+4.4\times 10^{-13}\tg$
   & fit by PAMS
   & MAP \\
H8) & {\bf H + H$^+$ $\rightarrow$ H$_2^+$ + $\gamma$}
   & ${\rm dex}[-19.38-1.523\log\tg+$ 
   &
   &  \\
   &
   & $1.118(\log\tg)^2-0.1269(\log\tg)^3]$
   & $1\leq\tg\leq 32000$, fit
   & RP, SBD \\
H9)& {\bf H$_2^+$ + $\gamma$ $\rightarrow$ H + H$^+$}
   & $2.0\times 10^1\tr^{1.59}\exp{\left(-\frac{82000}{\tr}\right)}$
   & $v=0$, fit
   & Du \\
   &
   & $1.63\times 10^7\exp{\left(-\frac{32400}{\tr}\right)}$
   & LTE, fit
   & Ar, St \\
H10) & {\bf H$_2^+$ + H $\rightarrow$ H$_2$ + H$^+$}
   & $6.4\times 10^{-10}$
   &
   & KAH \\
H11)& H$_2^+$ + e $\rightarrow$ 2H
   & $2.0\times 10^{-7}\tg^{-0.5}$
   & $v=0$, fit
   & SDGR \\
H12)& H$_2^+$ + $\gamma$ $\rightarrow$ 2H$^+$ + e
   & $9.0\times 10^1\tr^{1.48}\exp{\left(-\frac{335000}{\tr}\right)}$
   & fit
   & BO \\
H13)& H$_2^+$ + H$_2$ $\rightarrow$ H$_3^+$ + H
   & $2.0\times 10^{-9}$
   &
   & TH \\
H14)& H$_2^+$ + H $\rightarrow$ H$_3^+$ + $\gamma$
   & irrelevant 
   & 
   & KH \\
H15) & {\bf H$_2$ + H$^+$ $\rightarrow$ H$_2^+$ + H}
   & $3.0\times 10^{-10}\exp{\left(-\frac{21050}{\tg}\right)}$
   & $\tg\leq 10^4$, fit
   & \\
   &
   & $1.5\times 10^{-10}\exp{\left(-\frac{14000}{\tg}\right)}$
   & $\tg> 10^4$, fit
   & HMF \\
H16) & H$_2$ + e $\rightarrow$ H + H$^-$
   & $2.7\times 10^{-8}\tg^{-1.27}\exp{\left(-\frac{43000}{\tg}\right)}$
   & $v=0$, fit
   & SA \\
H17) & H$_2$ + e $\rightarrow$ 2H + e
   & $4.4\times 10^{-10}\tg^{0.35}\exp{\left(-\frac{102000}{\tg}\right)}$
   & fit by MD 
   & Co \\
H18)& H$_2$ + $\gamma$ $\rightarrow$ H$_2^+$ + e
   & $2.9\times 10^2\tr^{1.56}\exp{\left(-\frac{178500}{\tr}\right)}$
   & fit
   & OR \\
H19)& H$_3^+$ + H $\rightarrow$ H$_2^+$ + H$_2$
   & $7.7\times 10^{-9}\exp{\left(-\frac{17560}{\tg}\right)}$
   & fit
   & SMT  \\
H20)& H$_3^+$ + e $\rightarrow$ H$_2$ + H
   & $4.6\times 10^{-6}\tg^{-0.65}$
   &
   & Su  \\
H21)& H$_2$+H$^+$ $\rightarrow$ H$_3^+$ + $\gamma$
   & $1.0\times 10^{-16}$ 
   &
   & GH \\
H22)& H$_3^+$ + $\gamma$ $\rightarrow$ H$_2^+$ + H
   & irrelevant 
   & 
   & KH \\
\hline	
\end{tabular}
\vspace{1em}

\small{
Ar: Argyros (1974);
BO: Bates \& Opik (1968);
Co: Corrigan (1965);
DJ: de Jong (1972) ;
Du: Dunn (1968) ;
GH: Gerlich \& Horning (1992) ;
HMF: Holliday et al. (1971) ;
KAH: Karpas et al. (1979);
KH: Kulander \& Heller (1978);
LDZ: Launay et al. (1991);
MAP: Moseley et al. (1970) ;
MD: Mitchell \& Deveau (1983);
OR: ONeil \& Reinhardt (1978);
PAMS: Peterson et al. (1971);
Po: Poulaert et al. (1978);
RP: Ramaker \& Peek (1976);
SA: Schulz \& Asundi (1967);
SBD: Stancil et al. (1993);
SDGR: Schneider et al. (1994);
SMT: Sidhu et al. (1992);
St: Stancil (1994)
Su: Sundstr\"om et al. (1994);
TH: Theard \& Huntress (1974)
}
\end{flushleft}
\end{table*}

{\bf (H1)-(H2) H recombination}

The process of cosmological recombination has been analyzed in detail
in a number of studies, most recently by Jones \& Wyse (1985), Krolik
(1990), Grachev \& Dubrovich (1991), and Sasaki \& Takahara (1993). The
improved treatment of stimulated processes during recombination
followed by the last two goups has led to a more accurate estimate of
the electron abundance at freeze-out. Compared to the results of Jones
\& Wyse (1985) the residual ionization degree is a factor 2--3 lower,
depending on the cosmological model. Of course, the reduced number of
free electrons and protons has a direct effect on the chemical
evolution of the gas.  In this paper we have solved the full equation
for the time evolution of the electron abundance (Eq.~(1) of Sasaki
\& Takahara 1993) using the prescriptions for the rate of radiative
recombination given by Grachev \& Dubrovich (1991).  In practice, the
value for the rate of radiative transition from the continuum to the
excited states is given by the simple power-law fit $R_{c2}=8.76\times
10^{-11}(1+z)^{-0.58}$~cm$^3$~s$^{-1}$. The inverse rate from the
excited levels to the continuum is $R_{2c}=2.41\times 10^{15}
\tr^{1.5}\exp(-39472/\tr)R_{c2}$~s$^{-1}$.

\noindent
{\bf (H3) Radiative attachement of H}

We have fitted the data tabulated by de Jong (1972) (triangles), based
on the cross section of Ohmura \& Ohmura (1960).  As seen in
Fig.~1, our fit (solid line) agrees well with the expression used by
Abel et al. (1997) (dashed line).

\begin{figure*} 
\resizebox{\hsize}{!}{\includegraphics{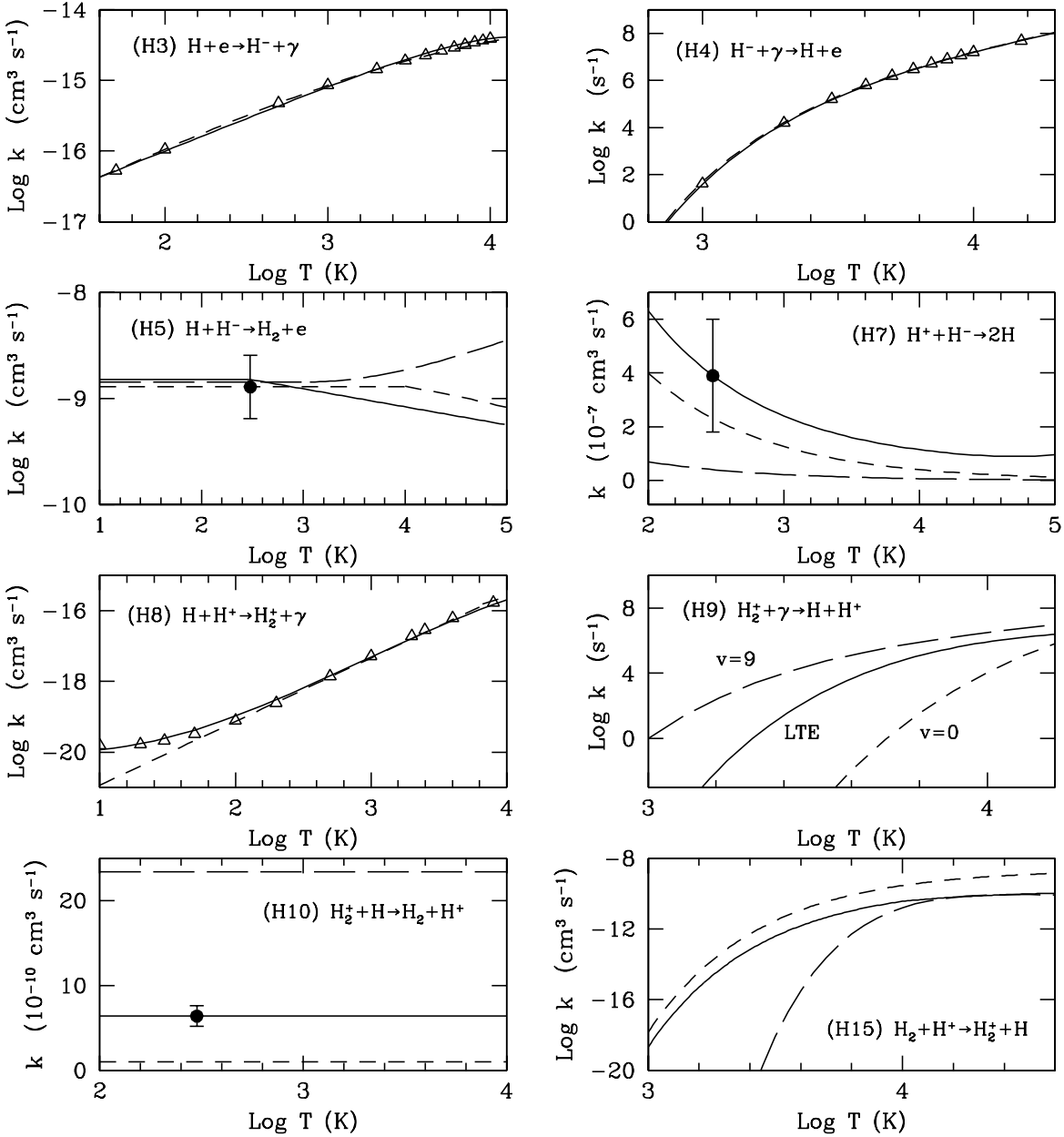}}
\caption[]{Rate coefficients for H$_2$ chemistry.}
\end{figure*}

\noindent
{\bf (H4) Radiative detachment of H$^-$}

The cross section for this reaction is known with great accuracy
(Wishart 1979).  We compare in Fig.~1 our fit (solid line) to the the
rate tabulated by de Jong (1972) (triangles).  The expression given by
Tegmark et al. (1997) is shown by a dashed line.

\noindent
{\bf (H5) Associative detachment of H and H$^-$}

There is substantial agreement between the value of this rate
coefficient (to within $\sim 10$\%) at temperatures below
$\sim$10$^3$~K and the experimental measurement at $T=300$~K performed
by Schmeltekopf et al. (1967).  At higher temperatures earlier results
by Browne \& Dalgarno (1969) indicated a slow rise in the value of the
rate, whereas the more recent calculations of Launay et al. (1991) show
a significant decline.  An uncertainty of about 50 \%, however, is
still present in the data due to the uncertainty on the interaction
potential. Our fit (solid line in Fig.~1) is based on the results by
Launay et al. (1991), and is in reasonable agreement with the fit of
Shapiro \& Kang (1987) to the data of Bieniek (1980). On the other
hand, Abel et al.  (1997) favor a rate increasing with temperature for
$T>1000$ K, assuming a rate based on the cross section of Browne \&
Dalgarno (1969).

\noindent
{\bf (H7) Mutual neutralization of H$^-$ and H$^+$}

This rate is still subject to large uncertainties at low temperatures
($T\la 2000$~K).  Our fit in Fig.~1 (solid line) is based on the cross
sections of Peterson et al. (1971) and Moseley et al. (1970) and the
experimental point at $T=300$~K of Moseley et al. (1970). At
temperatures below $\sim$10$^3$~K, we have extrapolated as a power-law
($\propto E^{-1}$) the cross section of Moseley et al. (1970). The rate
used by Abel et al. (1997) is the same as Dalgarno \& Lepp (1987) and it is
shown by the long-dashed line. Finally, the rate adopted by Shapiro
\& Kang (1987), taken from Duley \& Williams (1984), is shown by the
dashed line.

\noindent
{\bf (H8) Radiative association of H and H$^+$}

This reaction rate has been computed quantum-me\-cha\-ni\-cal\-ly by Ramaker \&
Peek (1976) and more recently by Stancil et al. (1993). The
two calculations agree to within 3\%  from $T=10$~K up to $T=10^6$~K.
In Fig.~1 we show the values tabulated by Ramaker \& Peek (1976)
(triangles), our polynomial fit (solid line), and the fit by Abel et
al. (1997) (dashed line) taken from Shapiro \& Kang (1987). Since this
reaction is important at large redshifts when the temperature is well
above $\sim$10$^3$~K, the departure of the fit at low $T$ is of little
consequence on the final H$_2$ abundances.

\noindent
{\bf (H9) Photodissociation of H$_2^+$} 

The evolution of H$_2^+$, and, to a large extent, that of H$_2$ depend
crucially on the rate of this reaction. Unfortunately, calculations of
the photodissociation cross section are available only for a sparse set
of vibrational levels (Dunn 1968, Argyros 1974). Total cross sections,
averaged over a LTE distribution of level populations, have been
computed by Argyros (1974) for $2500\;{\rm K}< T< 26000\;{\rm K}$ and
by Stancil (1994) for $3150\;{\rm K}<T< 25200\;{\rm K}$. The
theoretical calculations are in excellent agreement and reproduce
satisfactorly well the experimental results by von Busch \&
Dunn (1972). In Fig.~1 we show the LTE photodissociation rate (solid
line), compared with the rates for photodissociation from the $v=0$ and
$v=9$ levels (dashed and long-dashed lines, respectively). In our
standard model we adopt the LTE value, but we also show the sensitivity
of the results to other choices of the reaction rate.

\noindent
{\bf (H10) H$_2$ formation via H$_2^+$}

As in most other studies, we assume a rate independent of temperature,
corresponding to the measurement at $T=300$~K by Karpas et al.
(1979).  The Langevin rate is shown by the long-dashed line.  An
uncertainty of a factor of $\sim 6$ is still present in the literature
as indicated by the dashed line in Fig.~1 (Culhane \& McCray 1995). Such
a reduction only affects the final H$_2^+$ abundance by almost the same
factor, while leaves the H$_2$ abundance unchanged within a few
percent.

\noindent
{\bf (H15) Charge exchange of H$_2$ with H$^+$}

We have integrated the cross section measured by Holliday et al. (1971)
and our fit is shown by the solid line.  The dashed line shows the rate
used by Shapiro \& Kang (1987), based on a simple detailed balance from
the reverse reaction.  The rate computed by Abel et al. (1997) from the
calculations of Janev et al. (1987) shows a dramatic drop at $T<10^4$~K
(long dashed line in Fig.~1). However, Janev et al.'s (1987) data are
accurate only for $T$ higher than $\sim 10^4$ K, and should be
disregarded at lower temperatures.  In any case, as noted by Abel et
al. (1997), large differences in this range of temperatures do not
affect significantly the final abundances of H$_2$ and H$_2^+$.

\subsection{Deuterium chemistry}

\begin{table*}[t]
\caption{\sc Reaction rates for Deuterium species}
\vspace{1em}
\begin{flushleft}
\begin{tabular}{lllll}
\hline
   &  reaction   &   rate (cm$^3$~s$^{-1}$ or s$^{-1}$)   & notes &  reference \\
\hline
   &             &                            &       &            \\
D1) & {\bf D$^+$ + e $\rightarrow$ D + $\gamma$} & & see text & \\
D2) & {\bf D + $\gamma$ $\rightarrow$ D$^+$+ e} & & see text & \\
D3) & {\bf D + H$^+$ $\rightarrow$ D$^+$ + H}
   & $3.7\times 10^{-10}\tg^{0.28}\exp\left(-\frac{43}{\tg}\right)$
   & fit
   & WCD \\
D4) & {\bf D$^+$ + H $\rightarrow$ D + H$^+$}
   & $3.7\times 10^{-10}\tg^{0.28}$
   & fit
   & WCD \\
D5) & D + H $\rightarrow$ HD + $\gamma$
   & $1.0\times 10^{-25}$
   & estimate
   & LS \\
D6)& D + H$_2$ $\rightarrow$ H + HD 
   & $9.0\times 10^{-11}\exp\left(-\frac{3876}{\tg}\right)$ 
   & $\tg > 250$, fit
   & ZM  \\
D7) & HD$^+$ + H $\rightarrow$ H$^+$ + HD
   &
   & same as H$_2^+$ + H
   & KAH \\
D8) & {\bf D$^+$ + H$_2$ $\rightarrow$ H$^+$ + HD}
   & $2.1\times 10^{-9}$
   & 
   & SAA \\
D9)& HD + H $\rightarrow$ H$_2$ + D 
   & $3.2\times 10^{-11}\exp\left(-\frac{3624}{\tg}\right)$
   & $\tg > 200$
   & Sh \\
D10) & {\bf HD + H$^+$ $\rightarrow$ H$_2$ + D$^+$} 
   & $1.0\times 10^{-9}\exp\left(-\frac{464}{\tg}\right)$
   & 
   & SAA \\
D11)& HD + H$_3^+$ $\rightarrow$ H$_2$ + H$_2$D$^+$ 
   & $(2.1-0.4\log\tg)\times 10^{-9}$
   & fit 
   & ASa, MBH \\
D12) & D + H$^+$ $\rightarrow$ HD$^+$ + $\gamma$
   &
   & same as H + H$^+$ 
   & RP \\
D13) & D$^+$ + H $\rightarrow$ HD$^+$ + $\gamma$
   &
   & same as H + H$^+$ 
   & RP \\
D14) & HD$^+$ + $\gamma$ $\rightarrow$ H + D$^+$
   & 
   & same as H$_2^+ + \gamma$ 
   &  \\
D15) & HD$^+$ + $\gamma$ $\rightarrow$ H$^+$ + D
   & 
   & same as H$_2^+ + \gamma$ 
   &  \\
D16) & HD$^+$ + e $\rightarrow$ H + D
   & $7.2\times 10^{-8}\tg^{-1/2}$
   & fit 
   & St \\
D17) & HD$^+$ + H$_2$ $\rightarrow$ H$_2$D$^+$ + H
   & 
   & same as H$_2^+$ + H$_2$ 
   & TH \\
D18) & HD$^+$ + H$_2$ $\rightarrow$ H$_3^+$ + D
   & 
   & same as H$_2^+$ + H$_2$ 
   & TH \\
D19)& D + H$_3^+$ $\rightarrow$ H$_2$D$^+$ + H
   & $2.0\times 10^{-8}\tg^{-1}$
   & uncertain 
   & ASb \\
D20)& H$_2$D$^+$ + e $\rightarrow$ H + H + D
   & $1.0\times 10^{-6}\tg^{-1/2}\times 0.73$
   & 
   & Da, La \\
D21)& H$_2$D$^+$ + e $\rightarrow$ H$_2$ + D
   & $1.0\times 10^{-6}\tg^{-1/2}\times 0.07$
   & 
   & Da, La \\
D22)& H$_2$D$^+$ + e $\rightarrow$ HD + H
   & $1.0\times 10^{-6}\tg^{-1/2}\times 0.20$
   & 
   & Da, La \\
D23)& H$_2$D$^+$ + H$_2$ $\rightarrow$ H$_3^+$ + HD
   & $4.7\times 10^{-9}\exp\left(-\frac{215}{\tg}\right)$
   & $\tg \leq 10^2$
   &  \\
   & 
   & $5.5\times 10^{-10}$ 
   & $\tg > 10^2$
   & ASa, MBH, He \\
D24)& H$_2$D$^+$ + H $\rightarrow$ H$_3^+$ + D
   & $2.0\times 10^{-8}\tg^{-1}\exp\left(-\frac{632}{\tg}\right)$
   &
   & ASb \\
\hline
\end{tabular}
\vspace{1em}

\small{
ASa: Adams \& Smith (1981);
ASb: Adams \& Smith (1985);
Da: Datz et al. (1995); 
He: Herbst (1982); 
KAH: Karpas et al. (1979)
La: Larsson et al. (1996);
LS: Lepp \& Shull (1984);
MBH: Millar et al. (1989);
RP: Ramaker \& Peek (1976);
SAA: Smith et al. (1982);
Sh: Shavitt (1959);
St: Str\"omholm et al. (1995);
TH: Theard \& Huntress (1974);
WCD: Watson et al. (1978);
ZM: Zhang \& Miller (1989);
}
\end{flushleft}
\end{table*}

\noindent
{\bf (D1)-(D2) D recombination and ionization} 

These processes have been treated in the same way as hydrogen (see (H1) 
and (H2)).

\noindent
{\bf (D3)-(D4) Charge exchange of D and H}

These reactions control the evolution of D$^+$.  For (D3) we have
fitted the tabulated rate coefficient of Watson et al. (1978) which
shows a weak temperature dependence. The rate coefficient for the
reverse reaction (D4) is obtained by multiplying the direct one by
$\exp(43/T)$. The fit is shown in Fig.~2 together with the data by
Watson et al. (1978) (triangles). Puy et al. (1993) assume a constant
rate that differs from ours by a factor of 2 in the temperature
interval 10--1000~K.

\begin{figure*} 
\resizebox{\hsize}{!}{\includegraphics{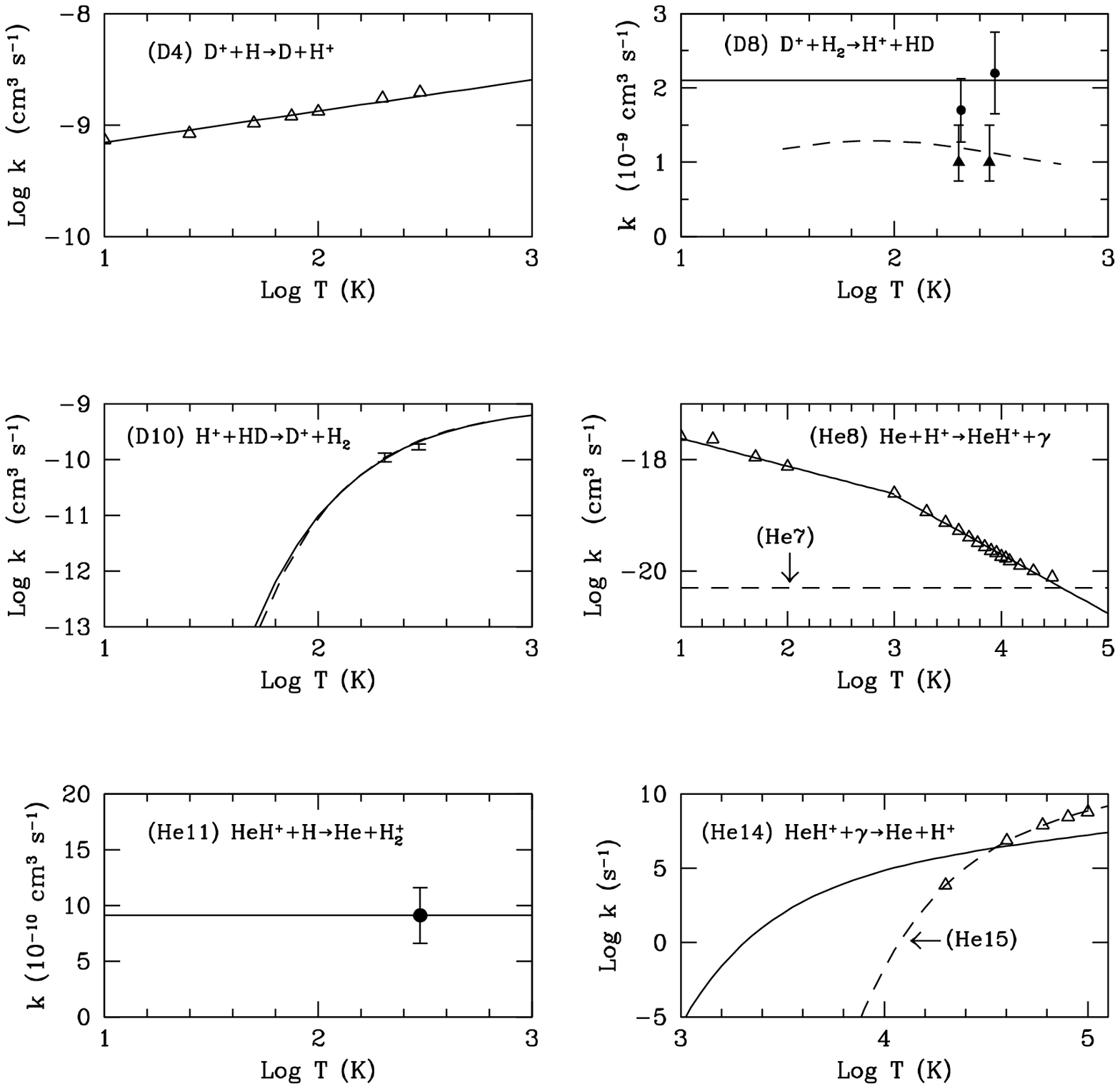}}
\caption[]{Rate coefficients for HD and HeH$^+$ chemistry.}
\end{figure*}

\noindent
{\bf (D8) Formation of HD via H$_2$}

We have used a rate equal to the Langevin collision rate constant
(solid line in Fig.~2). This rate is slightly bigger ($2.1\times
10^{-9}$ vs. $1.7\times 10^{-9}$) than the experimental result of Smith
et al. (1982) at $T=205$~K, but well in agreement at $T=300$~K (filled
circles).  These values are larger than the previous measurements by
Fehsenfeld et al. (1973) (filled triangles) and the theoretical
estimate by Gerlich (1982).

\noindent
{\bf (D10) Destruction of HD via H$^+$}

This is the reverse reaction of (D8) which is endoergic by 464 K. The
solid line in Fig.~2 is the rate given by Smith et al. (1982). The
dashed line is the theoretical result by Gerlich (1982).  The
data points at $T=205$ and 295~K are from Henchmann et al. (1981).

\subsection{Helium chemistry}

\begin{table*}[t]
\caption{\sc Reaction rates for Helium species}
\vspace{1em}
\begin{flushleft}
\begin{tabular}{lllll}
\hline
   &  reaction   &   rate (cm$^3$~s$^{-1}$ or s$^{-1}$)   & notes &  reference \\
\hline
   &             &                            &       &            \\
He1) & He$^{++}$ + e $\rightarrow$ He$^+$ + $\gamma$
   & $1.891\times 10^{-10}\left[\sqrt{\frac{\tg}{9.37}}
     \left(1+\sqrt{\frac{\tg}{9.37}}\right)^{0.2476}\right.\times$
   &
   &  \\
   &
   & $\left.\left(1+\sqrt{\frac{\tg}{2.774\times 10^6}}\right)^{1.7524}\right]^{-1}$
   & 
   & VF \\
He2) & He$^+$ + $\gamma$ $\rightarrow$ He$^{++}$ + e
   & $5.0\times 10^1\tr^{1.63}\exp\left(-\frac{590000}{\tr}\right)$
   & fit
   & Os \\
He3) & He$^+$ + e $\rightarrow$ He + $\gamma$
   & $3.294\times 10^{-11}\left[\sqrt{\frac{\tg}{15.54}}
     \left(1+\sqrt{\frac{\tg}{15.54}}\right)^{0.309}\right.\times$
   &
   &  \\
   &
   & $\left.\left(1+\sqrt{\frac{\tg}{3.676\times 10^7}}\right)^{1.691}\right]^{-1}$
   & 
   & VF \\
He4) & He + $\gamma$ $\rightarrow$ He$^+$ + e
   & $1.0\times 10^4\tr^{1.23}\exp\left(-\frac{280000}{\tr}\right)$
   & fit
   & Os \\
He5) & He + H$^+$ $\rightarrow$ He$^+$ + H
   & $4.0\times 10^{-37}\tg^{4.74}$
   & $\tg\ge 8000$, fit 
   & KLDD \\
He6) & He$^+$ + H $\rightarrow$ He + H$^+$
   & $3.7\times 10^{-25}\tg^{2.06}\times$
   &  
   & \\
   &
   & $\left[1+9.9\exp\left(-\frac{\tg}{2570}\right)\right]$
   & fit by KF $\tg>6000$
   & ZDKL \\
He7) & He + H$^+$ $\rightarrow$ HeH$^+$ + $\gamma$
   & $5.0\times 10^{-21}$
   & rad. ass.
   & RD, KLDD \\
He8) & {\bf He + H$^+$ $\rightarrow$ HeH$^+$ + $\gamma$}
   & $7.6\times 10^{-18}\tg^{-0.5}$
   & $\tg\leq 10^3$, 
   &  \\
   &
   & $3.45\times 10^{-16}\tg^{-1.06}$
   & $\tg> 10^3$ (inv. prediss.)
   & RD \\
He9) & He + H$_2^+$ $\rightarrow$ HeH$^+$ + H
   & $3.0\times 10^{-10}\exp\left(-\frac{6717}{\tg}\right)$
   &
   & Bl \\
He10) & He$^+$ + H $\rightarrow$ HeH$^+$ + $\gamma$
   & $1.6\times 10^{-14}\tg^{-0.33}$
   & $\tg\leq 4000$, fit
   & \\
   &  
   & $1.0\times 10^{-15}$
   & $\tg > 4000$, fit
   & ZD \\
He11) & {\bf HeH$^+$ + H $\rightarrow$ He + H$_2^+$}
   & $9.1\times 10^{-10}$
   & 
   & KAH \\
He12)& HeH$^+$ + e $\rightarrow$ He + H
   & $1.7\times 10^{-7}\tg^{-0.5}$
   & 
   & YM \\
He13)& HeH$^+$ + H$_2$ $\rightarrow$ H$_3^+$ + He
   & $1.3\times 10^{-9}$
   &
   & Or \\
He14)& {\bf HeH$^+$ + $\gamma$ $\rightarrow$ He + H$^+$}
   & $6.8\times 10^{-1}\tr^{1.5}\exp\left(-\frac{22750}{\tr}\right)$
   & fit
   & RD \\
He15)& HeH$^+$ + $\gamma$ $\rightarrow$ He$^+$ + H
   & $7.8\times 10^3\tr^{1.2}\exp\left(-\frac{240000}{\tr}\right)$
   & fit
   & RD \\ 
\hline
\end{tabular}
\vspace{1em}

\small{
Bl: Black (1978);
KAH: Karpas et al. (1979);
KLDD: Kimura et al. (1993);
Or: Orient (1977);
Os: Osterbrock (1989);
RD: Roberge \& Dalgarno (1982);
VF: Verner \& Ferland (1996);
YM: Yousif \& Mitchell (1989);
ZD: Zygelman \& Dalgarno (1990);
ZDKL: Zygelman et al. (1989);
}
\end{flushleft}
\end{table*}

{\bf (He8) Radiative association of He and H$^+$}

This rate of direct radiative association (He7) is quite small compared
to inverse rotational predissociation, as shown by Flower \& Roueff
(1979), Roberge \& Dalgarno (1982) and Kimura et al. (1993). The solid
line in Fig.~2 shows our fit to the photorates computed by Roberge \&
Dalgarno (1982) (triangles).  For comparison, the dashed line shows the
rate of direct radiative association (Roberge \& Dalgarno 1982).

\noindent
{\bf (He11) Destruction of HeH$^+$ by H}

The rate coefficient has been measured by Karpas et al. (1979)
at $T=300$~K (shown by a filled circle in Fig.~2).  We have used a
temperature-independent rate, which is equal to about one-half of the
Langevin collision rate constant.

\noindent
{\bf (He14) Photodissociation of HeH$^+$}

This rate has been computed from detailed balance of the direct
radiative association using a threshold energy of $1.96$~eV and the
rate coefficient of (He7). The result is shown by the solid line in
Fig.~2. For comparison we also show the rate for reaction (He15)
obtained by fitting (dashed line) the results of Roberge \&
Dalgarno (1982) (triangles).

\subsection{Lithium chemistry}

\begin{table*}[t]
\caption{\sc Reaction rates for Lithium species}
\vspace{1em}
\begin{flushleft}
\begin{tabular}{lllll}
\hline
   &  reaction   &   rate (cm$^3$~s$^{-1}$ or s$^{-1}$)   & notes &  reference \\
\hline
   &             &                            &       &            \\
Li1) & {\bf Li$^+$ + e $\rightarrow$ Li + $\gamma$ }
   & $1.036\times 10^{-11}[\sqrt{\tg/107.7}\times$
   &  
   & \\
   &  
   & $(1+\sqrt{\tg/107.7})^{0.612}\times$
   &
   & \\
   &
   & $(1+\sqrt{\tg/1.177\times 10^7})^{1.388}]^{-1}$ 
   & 
   & VF \\
Li2) & {\bf Li + $\gamma$ $\rightarrow$ Li$^+$ + e }
   & $1.3\times 10^3\tr^{1.45}\exp\left(-\frac{60500}{\tr}\right)$
   & det. bal. 
   &  \\
Li3) & Li$^+$ + H$^-$ $\rightarrow$ Li + H 
   & $6.3\times 10^{-6}\tg^{-0.5}-7.6\times 10^{-9}+$
   &  
   &  \\
   &   
   & $2.6\times 10^{-10}\tg^{0.5}+2.7\times 10^{-14}\tg$ 
   & fit
   & PHa \\
Li4) & {\bf Li$^-$ + H$^+$ $\rightarrow$ Li + H}
    &
    & same as Li$^+$ + H$^-$
    & PHa \\
Li5) & {\bf Li + e $\rightarrow$ Li$^-$ + $\gamma$ }
    & $6.1\times 10^{-17}\tg^{0.58}\exp{\left(-\frac{\tg}{17200}\right)}$ 
    & det. bal.
    & RBB \\
Li6) & {\bf Li$^-$ + $\gamma$ $\rightarrow$ Li + e  }
    & $1.8\times 10^2\tr^{1.4}\exp{\left(-\frac{8100}{\tr}\right)}$
    & fit
    & RBB \\
Li7) & Li + H$^+$ $\rightarrow$ Li$^+$ + H   
    & $2.5\times 10^{-40}\tg^{7.9}\exp{\left(-\frac{\tg}{1210}\right)}$ 
    & 
    & KDS \\
Li8) & Li + H$^+$ $\rightarrow$ Li$^+$ + H + $\gamma$      
    & $1.7\times 10^{-13}\tg^{-0.051}\exp{\left(-\frac{\tg}{282000}\right)}$ 
    & 
    & SZ \\
Li9) & {\bf Li$(^2S)$ + H $\rightarrow$ LiH + $\gamma$ }
   & $(5.6\times 10^{19}\tg^{-0.15}+7.2\times 10^{15}\tg^{1.21})^{-1}$ 
   & 
   & DKS, GGGa \\
Li10) & {\bf LiH + $\gamma$ $\rightarrow$ Li$(^2S)$ + H   }
   & $8.3\times 10^4\tr^{0.3}\exp\left(-\frac{29000}{\tr}\right)$ 
   & det. bal. 
   &  \\
Li11) & Li$(^2P)$ + H $\rightarrow$ LiH + $\gamma$ 
   & $2.0\times 10^{-16}\tg^{0.18}\exp{\left(-\frac{\tg}{5100}\right)}$  
   & $A^1\Sigma^+\rightarrow X^1\Sigma^+$
   & GGGb \\
Li12) & Li$(^2P)$ + H $\rightarrow$ LiH + $\gamma$ 
   & $1.9\times 10^{-14}\tg^{-0.34}$ 
   & $B^1\Pi\rightarrow X^1\Sigma^+$
   & GGGb \\
Li13) & Li + H$^-$ $\rightarrow$ LiH + e
    & $4.0\times 10^{-10}$ 
    & estimate 
    & SLD \\
Li14) & {\bf Li$^-$ + H $\rightarrow$ LiH + e  }
    & $4.0\times 10^{-10}$ 
    & estimate 
    & SLD \\
Li15) & LiH$^+$ + H $\rightarrow$ LiH + H$^+$
    & $1.0\times 10^{-11}\exp{\left(-\frac{67900}{\tg}\right)}$ 
    & estimate 
    & SLD \\
Li16) & LiH + H$^+$ $\rightarrow$ LiH$^+$ + H 
    & $1.0\times 10^{-9}$ 
    & estimate 
    & SLD \\
Li17) & {\bf LiH + H $\rightarrow$ Li + H$_2$  }
    & $2.0\times 10^{-11}$ 
    & estimate 
    & SLD \\
Li18) & {\bf Li$^+$ + H $\rightarrow$ LiH$^+$ + $\gamma$}
    & ${\rm dex}[-22.4+0.999\log\tg-0.351(\log\tg)^2]$
    & 
    & DKS, GGGa \\
Li19) & {\bf Li + H$^+$ $\rightarrow$ LiH$^+$ + $\gamma$}
    & $4.8\times 10^{-14}\tg^{-0.49}$
    &
    & DKS \\
Li20) & LiH + H$^+$ $\rightarrow$ LiH$^+$ + H
    & $1.0\times 10^{-9}$ 
    & estimate 
    & SLD \\
Li21) & LiH + H$^+$ $\rightarrow$ Li$^+$ + H$_2$ 
    & $1.0\times 10^{-9}$ 
    & estimate 
    & SLD \\
Li22) & {\bf LiH$^+$ + e $\rightarrow$ Li + H }
    & $3.8\times 10^{-7}\tg^{-0.47}$ 
    & estimate  
    & SLD \\
Li23) & LiH$^+$ + H $\rightarrow$ Li + H$_2^+$ 
    & $9.0\times 10^{-10}\exp{\left(-\frac{66400}{\tg}\right)}$ 
    & estimate 
    & SLD \\
Li24) & {\bf LiH$^+$ + H $\rightarrow$ Li$^+$ + H$_2$ }
    & $3.0\times 10^{-10}$ 
    & estimate 
    & SLD \\
Li25) & {\bf LiH$^+$ + $\gamma$ $\rightarrow$ Li$^+$ + H }
    & $7.0\times 10^2\exp{\left(-\frac{1900}{\tr}\right)}$
    & det. bal. 
    & DKS, GGGa \\
Li26) & LiH$^+$ + $\gamma$ $\rightarrow$ Li + H$^+$
    & $3.35\times 10^7 \tr\exp{\left(-\frac{97000}{\tr}\right)}$
    & det. bal. 
    & SLD \\
\hline
\end{tabular}
\vspace{1em}

\small{
DKS: Dalgarno et al. (1996);
GGGa: Gianturco \& Gori Giorgi (1996a);
GGGb: Gianturco \& Gori Giorgi (1996b);
KDS: Kimura et al. (1994);
PHa: Peart \& Hayton (1994);
RBB: Ramsbottom et al. (1994);
SLD: Stancil et al. (1996);
SZ: Stancil \& Zygelman (1996);
VF: Verner \& Ferland (1996)
}
\end{flushleft}
\end{table*}

\begin{figure*} 
\resizebox{\hsize}{!}{\includegraphics{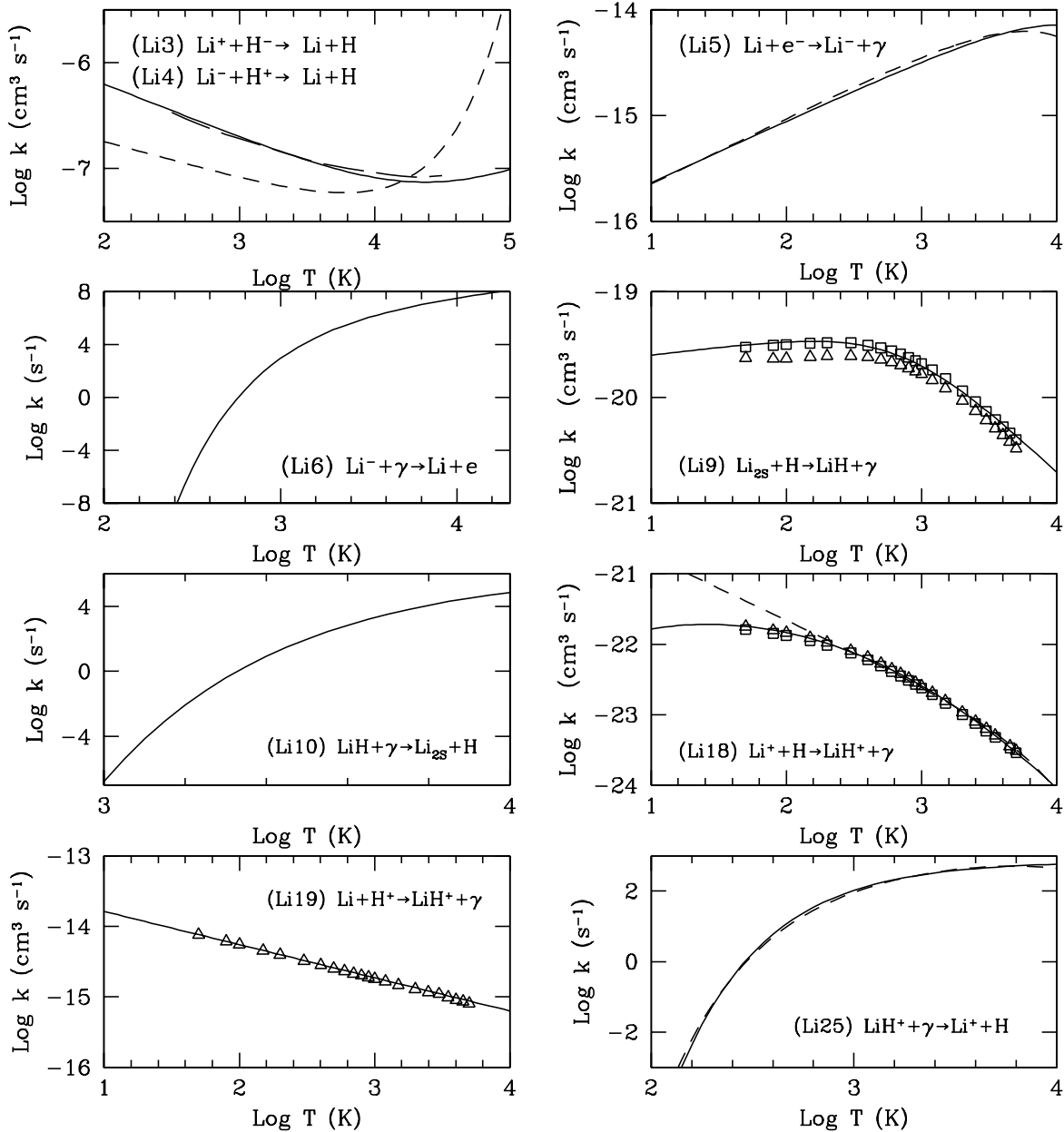}}
\caption[]{Rate coefficients for LiH chemistry.}
\end{figure*}

\noindent
{\bf (Li3)-(Li4) Mutual neutralizations of Li and H}

Following Stancil et al. (1996), we have adopted the same rate for
(Li3) and (Li4).  The solid line in Fig.~3 shows the rate obtained
using the cross section of Peart \& Hayton (1994). The long-dashed line
is the rate determined by Janev \& Radulovi\'{c} (1978). The dashed
line is the rate adopted by Stancil et al. (1996). 

\noindent
{\bf (Li5) Radiative attachment of Li and e$^-$}

We have used the cross section computed by Ramsbottom et al. (1994) for
the inverse process (Li6) and detailed balance arguments to compute the
rate shown in Fig.~3 by the solid line. The fit by Stancil et al.
(1996), corrected as in Stancil \& Dalgarno (1997), is shown by the
dashed line.

\noindent
{\bf (Li6) Photodetachment of Li$^-$}

The solid line shows the rate obtained from the
pho\-to\-de\-ta\-ch\-ment cross section evaluated by Ramsbottom et al. (1994).

\noindent
{\bf (Li9) Radiative association of LiH}

This rate has been recently determined quantum-me\-cha\-ni\-cal\-ly by
Dalgarno et al. (1996) (squares) and Gianturco \& Gori Giorgi
(1996a) (triangles). Our fit to the results of Dalgarno et al. (1996)
is shown by the solid line.

\noindent
{\bf (Li10) Photodissociation of LiH}

The solid line shows the analytical fit to the rate obtained by
detailed balance of reaction (Li9).

\noindent
{\bf (Li18) Radiative association of Li$^+$ and H}

We have fitted (solid line) the results of Dalgarno et al. (1996)
(squares).  The values obtained by Gianturco \& Gori Giorgi (1996a) are
shown by triangles. The dashed line represents the analytic fit by
Stancil et al. (1996).

\noindent
{\bf (Li19) Radiative association of Li and H$^+$ }

As in the case of (Li18), we have adopted the results of Dalgarno et al. (1996)
(triangles and solid line).

\noindent
{\bf (Li25) Photodissociation of LiH$^+$}

The dashed line shows the rate obtained by detailed balance applied to
the radiative association reaction (Li18) adopting the rate by Dalgarno
et al. (1996). Our fit is shown by the solid line.

\section{Evolutionary Models}

We consider now the chemical and thermal evolution of the pregalactic
gas in the framework of a Friedmann cosmological model. We start the
calculations at $z=10^4$ with all atomic species fully ionized and we
follow their recombination down to to $z=0$ assuming that no reionizing
events have taken place in the Universe.

In order to calculate the abundances of the 21 elements included in the
network, we have solved the set of coupled chemical rate equations of
the form
\beq
\label{abundances}
\frac{{\rm d}n_i}{{\rm d}t}=k_{\rm form} n_j n_k-k_{\rm dest}n_i+\ldots,
\enq
where $k_{\rm form}$ and $k_{\rm dest}$ are the formation and destruction 
reaction rates discussed in Sect.~2, and $n_i$ is the number density of 
the reactant species $i$.

The temperature of the CBR is given by $\tr=T_0(1+z)$ with
$T_0=2.726$~K (Mather et al. 1994). The evolution of the gas
temperature $\tg$ is governed by the equation (see e.g. Puy et al.
1993; Palla et al. 1995)
\beq
\label{tempe}
\frac{{\rm d}T_{\rm g}}{{\rm d}t}=-2T_{\rm g}\frac{\dot{R}}{R}+
\frac{2}{3kn}[(\Gamma-\Lambda)_{\rm Compton}+(\Gamma-\Lambda)_{\rm mol}].
\enq
The first term represents the adiabatic cooling associated with the
expansion of the Universe, $R$ being the scale factor.  The other two
terms represent respectively the net transfer of energy from the CBR
to the gas (per unit time and unit volume)
via Compton scattering of CBR photons on electrons
\beq
(\Gamma-\Lambda)_{\rm Compton}=\frac{4k\sigma_{\rm T}aT_{\rm r}^4
(T_{\rm r}-T_{\rm g})}{m_{\rm e}c}n_{\rm e},
\enq
and via excitation and de-excitation of molecular transitions
\beq
(\Gamma-\Lambda)_{\rm mol}=\sum_k n_k\sum_{i>j}(x_iC_{ij}-x_jC_{ji})h\nu_{ij},
\enq
where $C_{ij}$ and $C_{ji}$ are the collisional excitation and
de-excitation coefficients and $x_{i}$ are the fractional level
populations.  For the molecular heating and cooling of the gas we have
considered the contribution of H$_2$, HD and LiH. A full discussion of
the molecular parameters is given in the Appendix, where we present
analytical fits and plots of the cooling functions in the temperature
range $10\;{\rm K}\leq \tg \leq 10^4\;{\rm K}$. Given the large range
of validity, the cooling functions can be used in a variety of
cosmological applications.

The energy transfer function $(\Gamma-\Lambda)_{\rm mol}$ can become an
effective heating (cooling) source for the gas if the rate of
collisional de-excitation of the roto-vibrational molecular levels is
faster (slower) than their radiative decay (see Khersonskii 1986, Puy
et al. 1993). In general, however, the contribution of molecules to the
heating/cooling of the gas is very small.

The chemical network is completed by the equation for the redshift 
\beq
\frac{{\rm d}t}{{\rm d}z}=-\frac{1}{H_0(1+z)^2\sqrt{1+\Omega_0z}}, 
\enq
where $H_0=100\;h$~km~s$^{-1}$~Mpc$^{-1}$ is the Hubble constant 
and $\Omega_0$ the closure parameter.

The density $n(z)$ of baryons at redshift $z$ is 
\beq 
n(z)=\Omega_{\rm b}n_{\rm cr}(1+z)^3, 
\enq 
where $n_{\rm cr}=9.19\times 10^{-6} h^2$~cm$^{-3}$ is the critical
density, and $\Omega_b$ is the ratio of the total to the critical
density. This quantity is related to the baryon-to-photon ratio $\eta$
by $\Omega_b=3.66\times 10^{-3}\eta_{10}h^{-2}$, with
$\eta_{10}=10^{10}\eta$.

\section{Results}

\subsection{The standard model}

Our standard model is characterized by a Hubble constant $h=0.67$ (van
den Bergh 1989), a closure parameter $\Omega_0=1$, and a
baryon-to-photon ratio $\eta_{10}=4.5$ (Galli et al. 1995).  The
initial fractional abundance of H, D, He and Li are taken from the
standard big bang nucleosynthesis model of Smith et al. (1993).

The results are shown in Fig.~4, separately for each atomic element.
The residual ionization fraction at $z=1$ is $[{\rm e}/{\rm
H}]=3.02\times 10^{-4}$. Comparison with Model~III of SLD which has the
same values of the cosmological parameters (see Table 5) shows that the
freeze-out ionization fraction is a factor of 2 smaller in our case,
owing to the different treatment of H recombination.

\begin{table}[t]
\caption{\sc Ionization fraction at $z=10$}
\vspace{1em}
\begin{flushleft}
\begin{tabular}{llllll}
\hline
$H_0$ & $\Omega_0$ & $\Omega_b$ & $T_0$ & [e/H] & [e/H] \\
 & & & (K) & (this work) & (other) \\
\hline
67 & 1 & 0.0367 & 2.726 & 3.3(-4) & 6.5(-4) [SLD-III] \\
50 & 0.1 & 0.1 & 2.7 & 5.1(-5) & 5.0(-5)  [B] \\
50 & 1 & 0.1 & 2.726 & 1.7(-4) & 3.2(-4)  [SLD-I] \\
100 & 1 & 0.1 & 2.7 & 8.3(-5) & 7.8(-5)  [LB] \\
\hline	
\end{tabular}
\vspace{1em}

\small{
B: Black (1991);
LB: Latter \& Black (1991);
SLD-I, SLD-III: Models I and III of Stancil et al. (1996)
}
\end{flushleft}
\end{table}

\begin{figure*} 
\resizebox{\hsize}{!}{\includegraphics{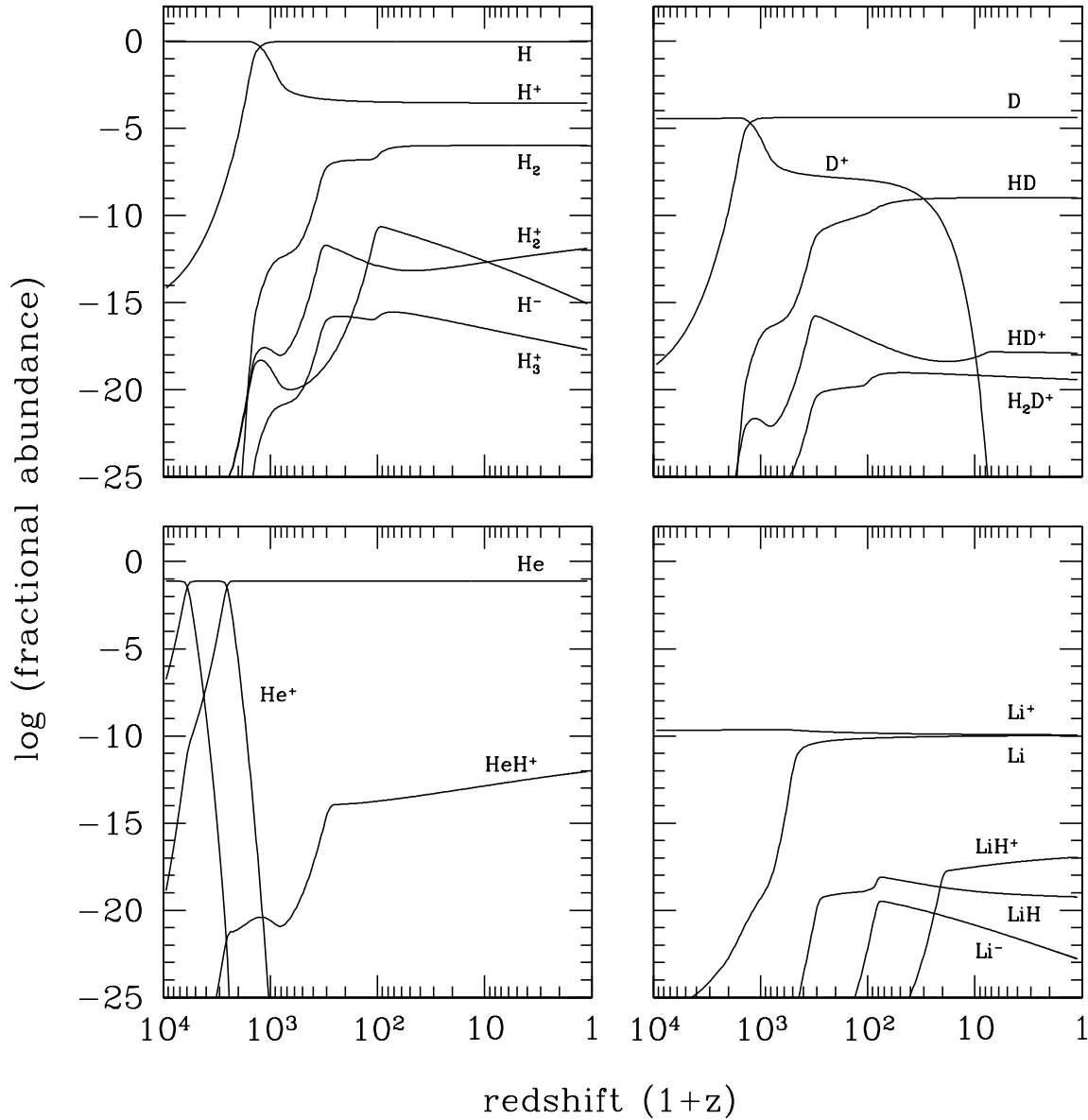}}
\caption[]{The evolution of all the chemical species considered in the standard
model as a function of redshift. The four panels show the results for 
H (upper left), D (upper right), He (lower left), and Li (lower right).}
\end{figure*}

The evolution of the abundance of H$_2$ follows the well known
behaviour (e.g. Lepp \& Shull 1983, Black 1991) where the initial steep
rise is determined by the H$_2^+$ channel (reactions H9 and H10)
followed by a small contribution from H$^-$ at $z\simeq 100$ (reactions
H3, H4 and H5). The freeze-out value of H$_2$ is [H$_2$/H] $=1.1\times
10^{-6}$, a value somewhat lower than found in similar studies, as a
consequence of the net reduction of the ionization fraction.

Fig.~4 shows that all the ionic species reach very low fractional
abundances.  In particular, the steady drop of H$^-$ at $z\la 100$ is
determined by mutual neutralization with H$^+$ (reaction H7).  As
discussed before, the rate for this reaction is still uncertain, but
the net effect on the final abundance of H$_2$ is quite small. The
reason is that the rates differ only at low temperatures (below $\sim
100$ K) when the H$_2$ formation via the H$^-$ channel is already
completed.  In fact, adopting the Dalgarno \& Lepp (1987) rate for
reaction (H7) leads to a 10\% difference in the H$_2$ abundance.  On
the contrary, the abundance of H$^-$ is quite affected by the choice of
the rate of (H7):  our final H$^-$ abundance at $z=1$ is three times
smaller than that obtained using Dalgarno \& Lepp (1987).

The formation of H$_2^+$ is controlled by radiative association (H8)
and photodissociation (H9) at $z\geq 100$, whereas at lower redshifts
there is a contribution from HeH$^+$ (reaction He11) that explains the
gentle rise up to [H$_2^+$/H$_2$] $\simeq 10^{-6}$.  An additional
channel for H$_2^+$ formation is the associative ionization reaction
${\rm H}+{\rm H}(n=2)\rightarrow {\rm H}_2^+ + {\rm e}$ (Rawlings et
al.~1993; see also Latter \& Black~1991 for H$_2$ formation). The rate
coefficient for this process is $\sim 4$ orders of magnitude faster
than (H8) but the extremely low fractional population of the $n=2$
level of H ($\ll 10^{-13}$) makes associative ionization not
competitive with radiative association.  The evolution of H$_2^+$ and,
to a large extent, the abundance of H$_2$ depend crucially on the
adopted photodissociation rate. This is clearly shown in Fig. 5 where
the results obtained with various choices of the rate discussed in
Sect. 2.1 are compared with the results of the standard model (shown by
the solid line).  Photodissociation of H$_2^+$ from $v=0$ (dashed line)
would result in an enhancement of a factor $\sim 200$ of the final
H$_2$ abundance, whereas photodissociation from $v=9$ (long-dashed
line), would delay the redshift of formation of H$_2^+$ from $z\simeq
10^3$ to $z\simeq 300$ and of H$_2$ from $z\simeq 600$ to $z\simeq
250$.  However, the asymptotic abundances of both species are left
unchanged.

We believe that the uncertainty in the redshift evolution of H$_2$ is
not as large as that shown in Fig.~5, because photodissociation from
$v=0$ is extremely unlikely to occur at $z\simeq 10^3$. The actual
behaviour of the H$_2$ abundance should be intermediate between the two
curves in Fig.~5 corresponding to photodissociation from $v=9$ and
LTE.  The conditions for a distribution of level populations in LTE may
be marginally satisfied, considering the fact that the radiative decay
rates (Posen et al. 1983) are much faster than the collisional
excitation rates at the typical densities, but we know from the work of
Ramaker \& Peek (1976) that the reverse reaction (H8) forms H$_2^+$
preferentially in excited states.  In such a case, photodissociation
would take place from high vibrational levels and the LTE rate may
underestimate somehow the actual rate. Thus, we conclude that the
uncertainty is limited to the redshift of formation of H$_2$ molecules
and not to their final abundance.

\begin{figure} 
\resizebox{\hsize}{!}{\includegraphics{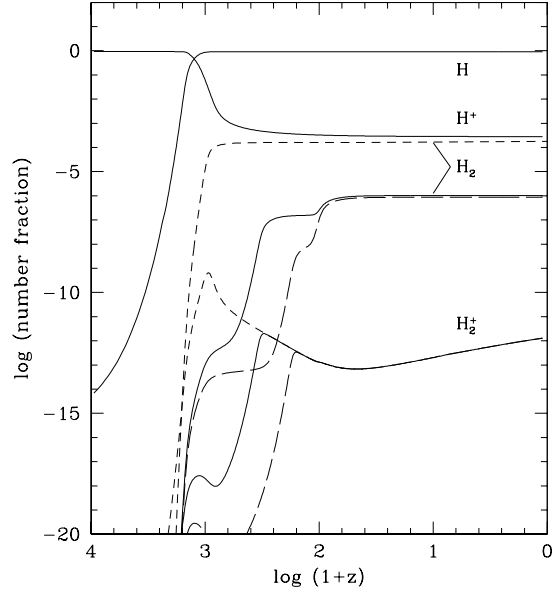}}
\caption[]{Effects of varying the photodissociation rate of H$_2^+$
on the evolution of H$_2$ and H$_2^+$. The solid curves are for the 
standard models, whereas the results obtained with photodissociation
from the $v=0$ and $v=9$ vibrational levels of H$_2^+$ are shown by the 
{\em dashed} and {\em long dashed} lines, respectively.}
\end{figure}

As for deuterium, Fig.~4 shows that the evolution of D$^+$ is very
sensitive to the charge exchange reactions (D3)--(D4). At $z<40$, the
abundance of D$^+$ drops precipitously due to the exponential factor in
the reaction rate (D3). The only molecule formed in significant amount
is HD, whose evolution with redshift follows closely that of H$_2$.
However, its abundance at freeze-out is [HD/H$_2$] $\simeq 10^{-3}$.
More important, the asymptotic value of [HD/H$_2$] becomes about
$10^{-3}$ with an enhancement factor of $\sim$30 from the initial [D/H]
abundance. This is due to the large fractionation implied by reaction
(D8).  The behaviour of HD$^+$ follows that of H$_2^+$ up to $z\simeq
100$, while at lower redshifts its evolution is dominated by reaction
(D12). The flattening at $z<10$ reflects the behaviour of the reaction
rate at temperatures below $\sim 20$~K.

The case of \hhdp is of interest after the suggestion by Dubrovich
(1993) and Dubrovich \& Lipovka (1995) that in the early universe this
molecule could reach significant abundances (up to [H$_2$D$^+$/H$_2$]
$=10^{-5}$) and that spectral features associated with the
rovibrational transitions might be detectable in the spectrum of the
CBR.  Such a high abundance would result from the complete conversion
of H$_3^+$ into H$_2$D$^+$ because of deuterium fractionation at low
temperatures ($T < 50$~K). However, it is very unlikely that this
mechanism could work in the conditions of the primordial gas, since the
abundance of H$_3^+$ ions is quite low due to the absence of ionizing
sources (see Fig.~4). In present-day molecular clouds, searches for the
rotational transitions of H$_2$D$^+$ have yielded a fractional
abundance of $3\times 10^{-11}$ (Boreiko \& Betz 1993).  The results
shown in Fig.~4 indicate that the final abundance of \hhdp is extremely
small [\hhdp/H] $\simeq 5\times 10^{-14}$, and [\hhdp/H$_3^+$] $\simeq
10^{-2}$.  It is clear from our models that the presence of H$_2$D$^+$
molecules can hardly affect the shape of the CBR (in order to see
detectable effects on the CBR, Dubrovich \& Lipovka (1993) assume an
abundance of 10$^{-8}$).  On the other hand, the isotopic ratio is
greatly enhanced with respect to the primordial abundance [D/H]
$=4\times 10^{-5}$, due to fractionation effects.

According to the results shown in Fig.~4, the main molecular species
containing helium is HeH$^+$, formed by the radiative association of He
and H$^+$. As emphasized by Lepp \& Shull (1984), this reaction is
slower than the usual formation mode via association of He$^+$ and H,
but since the abundance of He$^+$ is quite small, reaction (He8) takes
over. The HeH$^+$ ions are removed by CBR photons and at $z\la 250$
also by collisions with H atoms to reform H$_2^+$ (reaction He11). This
explains the abrupt change in the slope of the curve of HeH$^+$ shown
in Fig.~4. The final abundance is small, [HeH$^+$/H] $\sim 6\times
10^{-13}$ at $z=1$.

Finally, the chemistry of lithium is complicated, but the molecular
abundances are indeed very small. The more abundant complex is LiH$^+$
whose formation is controlled by the radiative association of Li$^+$
and H (see Dalgarno \& Lepp 1987). As shown in Fig. 4, lithium remains
more ionized at low redshifts and this explains why LiH is less
abundant than LiH$^+$. The final abundance of LiH$^+$ is $\simeq
10^{-17}$ at $z=1$.

\subsection{Dependence on the cosmological parameters}

The final values of the molecular abundances depend on the choice of
the cosmological parameters. Each model is in fact specified by three
parameters $\eta_{10}$, $\Omega_0$ and $h$. The observational (and
theoretical) uncertainties associated with each of them are still
rather large. We have selected as standard model that that better fits
the constraints imposed by the observations of the abundances of the
light elements. To gauge the effects that a change of $\eta_{10}$
induces on the chemistry network, we have run a model with
$\eta_{10}$=8.0, the maximum value still compatible with the
constraints on standard big bang nucleosynthesis given by observations
of lithium in Pop-II stars (Bonifacio \& Molaro 1997).  The results are
given in Table 6 for a direct comparison with the standard case. Note
that the abundance of H$_2$ is the same in the two models, making this
molecule a poor diagnostic of cosmological models. Lithium molecules
are quite sensitive to $\eta_{10}$, showing an increase of a factor of
3--4. For the other species, however, a higher value of $\eta_{10}$
implies a lower final abundance as a result of the lower initial value
of each element predicted by the standard big bang nucleosynthesis
which compensates for the higher total baryon density.  As shown by
Palla et al. (1995), larger variations of the molecular abundances than
those shown in Table 6 can be expected for more drastic variations of
the other cosmological parameters: both LiH and HD can vary by 2 or 3
orders of magnitudes by allowing changes to $\Omega_0$ and $h$ of a
factor of 5 and 2, respectively. However, current observations indicate
that a high value of the Hubble constant is unlikely, and that a
universe with $\Omega_0<1$, as in the case of a nonzero value of the
cosmological constant, is less favored by theoretical models.

\begin{table*}[t]
\caption{\sc Sensitivity of abundances to $\eta_{10}$}
\vspace{1em}
\begin{flushleft}
\begin{tabular}{llllllllllllll}
\hline
 & & & \multicolumn{4}{c}{$z=2000$} & & \multicolumn{6}{c}{$z=1$} \\
\cline{4-7} \cline{9-14}
$\eta_{10}$ & $X$ & $Y$ & H & He & D & Li & & [e/H] & [H$_2$/H] & [HD/H] & [HeH$^+$/H] & [LiH$^+$/H] & [LiH/H] \\
\hline
4.5 & 0.757 & 0.243 & 0.926 & 7.4(-2) & 4.0(-5) & 2.2(-10) & & 3.0(-4) & 1.1(-6) & 1.2(-9) & 6.2(-13) & 9.4(-18) & 7.1(-20) \\ 
8.0 & 0.751 & 0.249 & 0.923 & 7.6(-2) & 1.7(-5) & 7.0(-10) & & 1.7(-4) & 1.1(-6) & 5.0(-10) & 4.3(-13) & 2.5(-17) & 2.3(-19) \\
\hline	
\end{tabular}
\end{flushleft}
\end{table*}

\section{The minimal model}

The chemical network discussed in Sect. 2 consists of 87 reactions,
69 collisional and 18 radiative.  Although the integration of such
system does not present particular difficulties (apart for the
intrinsic stiffness of the rate equations) or require exceedingly long
computer times, for a better understanding of the chemistry of the
primordial gas it is more convenient to reduce the reactions to only
those that are essential to accurately model the formation/destruction
of each molecular species. Such a reduced system has been called the
{\em minimal model}. Abel et al. (1997) have devised a reduced network
in their discussion of the non-equilibrium effects and chemical
dynamics of the primordial gas. However, their application differs from
ours, since they are mainly interested in the formation of H$_2$ in the
post-shock layer of a cosmological pancake and consider only
collisional processes.

Our minimum model consists of the 33 reactions indicated in the diagram
shown in Fig.~6. The main reactions are listed in bold face in Tables~1
to 4. Notice that in order to describe fully the H$_2$ chemistry only
11 reactions out of 22 are needed. The formation of H$_2$ involves two
reaction sequences with H$^-$ and H$_2^+$ ions whose abundance is
restricted by photodestruction processes. However, for H$^-$ a
competing destruction channel is the mutual neutralization with H$^+$
ions which contributes $\sim 20$--40\% to the total rate at redshifts
$z\la 100$.  The H$_2$ system would be closed if not for the external
source of H$_2^+$ ions coming from the destruction of HeH$^+$ by H
atoms (reaction He11) at redshifts $z\la 250$.  It is important to
consider this path since it modifies substantially the evolution of
H$_2^+$ (see Fig.~4), although it does not affect the final abundance
of H$_2$ molecules. As we have already noted, in the absence of this
channel the final abundance of H$_2^+$ would be of the same order as
that of H$_3^+$, because of the importance of reaction (H19).

The chemistry of deuterium is rather simple and only 6 reactions out of
24 need to be considered. The molecule HD is formed rapidly by reaction
(D8) which involves ionized deuterium. Thus, it is important to
consider all the reactions that determine the ionization balance.  The
main destruction route is by reaction (D10), while other reactions
involving H$_3^+$ (D11) and H (D9) are much less important.

Finally, the chemistry of Li is more complex than the rest and even
the reduced network contains a large number of reactions (14 out of
26). This is mainly due to the fact that lithium remains partly ionized
and the routes to the formation of LiH and LiH$^+$ compete effectively
with each other. Note also that at redshifts $z\la 100$, the main route
to LiH formation is from associative detachment of Li$^-$ ions with H.
This process gradually removes Li$^-$ and the abundance of LiH reaches
a constant value.

\begin{figure} 
\resizebox{\hsize}{!}{\includegraphics{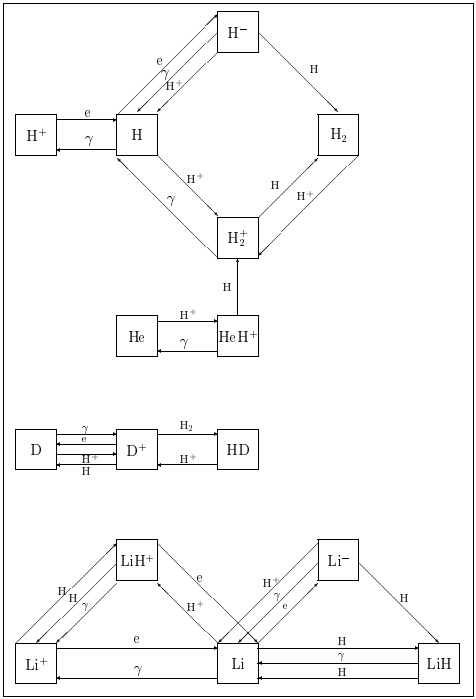}}
\caption[]{The minimal model for the main chemical species}
\end{figure}

\section{Comparison with previous work}

In order to make accurate comparisons with the results obtained in
previous studies, we have run models with the same reaction rates
adopted in our standard case, but differing cosmological parameters to
reproduce exactly the choices of the various authors listed in
Table~7.

\begin{table}[t]
\caption{\sc Values of the cosmological parameters}
\vspace{1em}
\begin{flushleft}
\begin{tabular}{lllll}
\hline
Model & $h$ & $\Omega_0$ & $\Omega_b$ & $T_0$ \\
      &     &            &            & (K)   \\
\hline
PGS   & 67    & 1          & 0.0367     & 2.726 \\
LS    & 50    & 0.1        & 0.1        & 3     \\
PALLP & 50    & 1          & 0.1        & 2.7   \\
B     & 50    & 0.1        & 0.1        & 2.7   \\
GS    & 75    & 1          & 0.02       & 2.7   \\
\hline	
\end{tabular}
\vspace{1em}

\small{
PGS: Palla et al. (1995);
LS: Lepp \& Shull (1983);
PALLP: Puy et al. (1993);
B: Black (1991);
GS: Giroux \& Shapiro (1996)
}
\end{flushleft}
\end{table}

The results are
shown in Fig.~7, where we compare the predictions of our model (solid
lines) with those of Palla et al. (1995) (upper left panel),
Lepp \& Shull (1984) (upper right panel), Puy et al. (1993) (lower
left panel), and Black (1991) (lower right panel), all indicated by
dashed lines. A common feature in the first three cases is a
dramatic reduction of the abundance of LiH by 7--9 orders of magnitude,
due to the replacement of the semiclassical estimate of the rate of
radiative association (Lepp \& Shull 1984) with the quantal
calculations by Dalgarno et al. (1996) and Gianturco \& Gori
Giorgi (1996a).

With respect to our earlier study of primordial chemistry, the more
accurate treatment of H recombination results in a decreased residual
electron fraction by a factor 2--3 (see also Table~5). This, in
turn, has a direct impact on the H$_2$ and HD abundances, which are
lower by a factor $\sim 5$. Otherwise, the evolution with redshift of
these two species follows the same behaviour.

Similar considerations apply to the comparison with the results
obtained earlier by Lepp \& Shull (1984) for H, H$^+$ and H$_2$.
However, it is worth noticing that in their model the formation of HD
occurs through the analogues of the H$^-$ and H$_2^+$ channels, at
rates diminished by the cosmological D/H ratio, plus a negligible
contribution from direct radiative association of H and D. In contrast,
we have found that the main reaction responsible for the formation of
HD is the isotope exchange reaction (D8) which has no counterpart in
the hydrogen chemical network. Therefore, despite their higher residual
electron fraction, their asymptotic abundance of HD is underestimated
by about one order of magnitude.

Puy et al. (1993) obtained abundances of H$_2$ and HD about one order
of magnitude larger than ours, owing to their unrealistic choice of the
photodissociation rate (H9) by von Bush \& Dunn (1968), valid for a LTE
distribution of level populations at $T=100$~C.  The higher abundance of
H$_2^+$ results in a higher abundance of H$_2$ formed through this
channel, and, in turn, to a correspondingly larger abundance of HD
through reaction (D8).

Finally, the results of Black (1991), Latter \& Black (1991), and
Giroux \& Shapiro~(1996) allow a direct comparison of the evolution of
some less abundant species like H$^-$, H$_2^+$ and HeH$^+$. The
agreement with our results and those by Black (1991) is extremely good:
the abundance of H, H$^+$ and H$_2$ agree within less than a factor 1.5
from $z=10^3$ to $z=10$, while the abundances of H$^-$, H$_2^+$,
although differing by a factor $\sim 3$, follow remarkably the same
behaviour with redshift. Only HeH$^+$ shows a marked (but
unconsequential) difference for $z>300$, probably due to a different
choice of the photodissociation rate (He14). On the other hand, the
comparison with Giroux \& Shapiro~(1996) shows a different behaviour of
H$_2^+$ at $z\la 100$ resulting from the neglect of the HeH$^+$ channel
for H$_2^+$ formation.

\begin{figure*} 
\resizebox{\hsize}{!}{\includegraphics{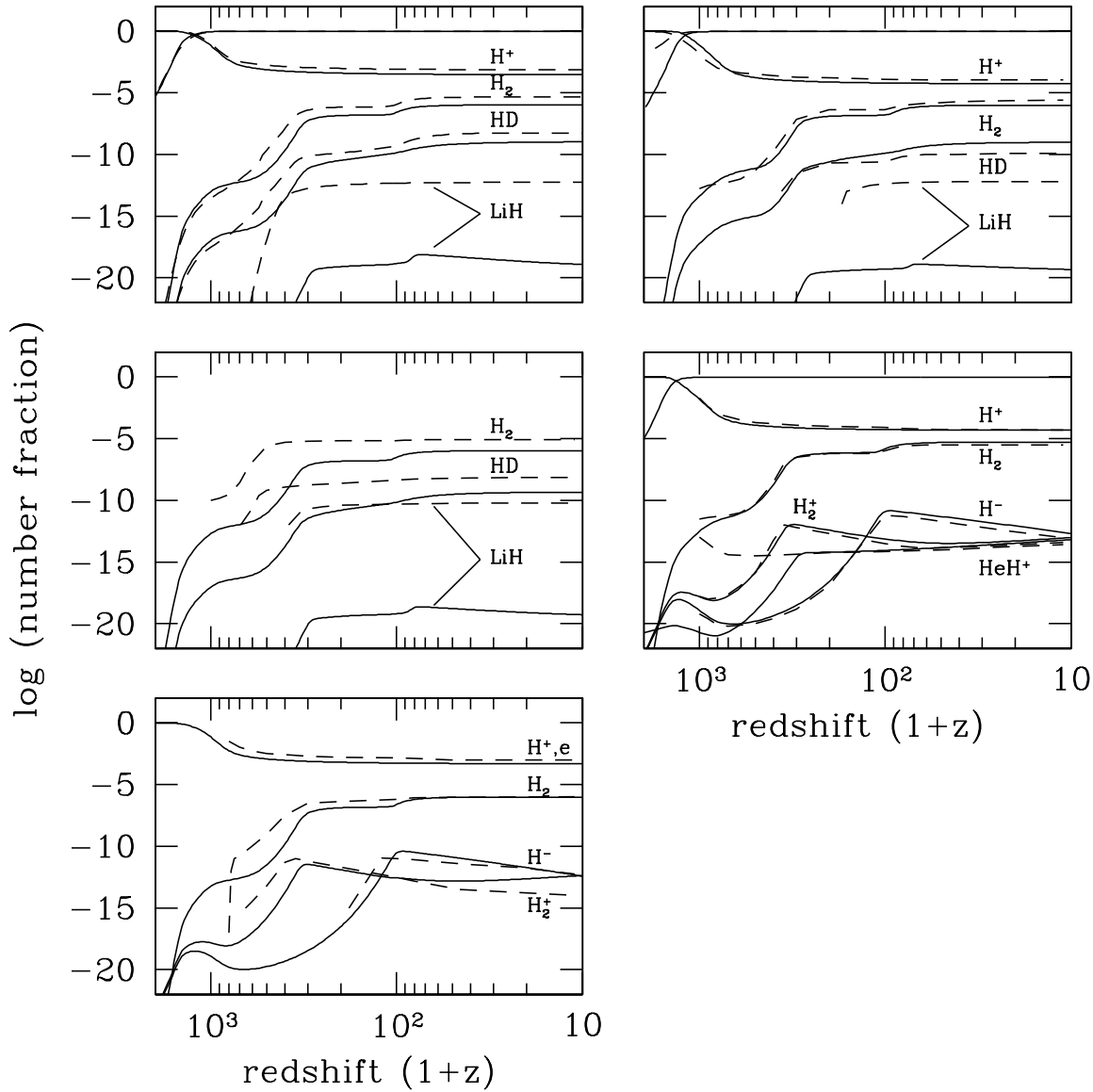}}
\caption[]{Comparison between the results of the standard model (solid lines) 
and those of
Palla et al. (1995) ({\em upper left}), Lepp \& Shull  (1984)
({\em upper right}), Puy et al. (1993) ({\em middle left}), 
Black (1991) ({\em middle right}), and Giroux \& Shapiro~(1996) ({\em lower left}),
all indicated by dashed lines.} 
\end{figure*}

\section{Conclusions}

The main results of the present study can be summarised as follows:

1) We have followed the chemical evolution of the primordial gas after
recombination by computing the abundances of 21 species, 12 atomic and 9
molecular, by using a complete set of reaction rates for collisional and
radiative processes. The rates which are critical for a correct estimate
of the final molecular abundances have been analysed and compared in detail.

2) One of the major improvements of this work is the use of a better
treatment of H recombination that leads to a reduction of a factor 2--3
in the abundance of electrons and H$^+$ at freeze-out, with respect to
previous studies. The lower residual ionization has a negative effect
on the chemistry of the primordial gas in which electrons and protons
act as catalysts in the formation of the first molecules.

3) In the standard model ($h=0.67$, $\eta_{10}=4.5$, $\Omega_0=1$ and
[D/H] $=4.3\times 10^{-5}$), the residual fractional ionization at $z=1$
is $[{\rm e/H}]=3.02\times 10^{-4}$, and the main molecular species have
fractional abundances $[{\rm H}_2/{\rm H}]=1.1\times 10^{-6}$, $[{\rm
HD/H}_2]=1.1\times 10^{-3}$, $[{\rm HeH}^+/{\rm H}]=6.2\times 10^{-13}$,
$[{\rm LiH}^+/{\rm H}]=9.4\times 10^{-18}$ and $[{\rm
LiH/LiH}^+]=7.6\times 10^{-3}$.

4) As for molecular hydrogen, its final abundance does not depend on
the model parameters, making this molecule a poor diagnostic of
cosmological scenarios. The largest uncertainty resides in the accurate
knowlwedge of the photodissociation rate of H$_2^+$. A detailed treatment
of the reaction kinetics of this reaction would be required.

5) We have presented a minimal model consisting of 11 reactions for
H$_2$, 6 for HD, 3 for HeH$^+$ and 14 for LiH which reproduces with
excellent accuracy the results of the full chemical network,
regardless of the choice of the cosmological parameters.

6) Finally, we have computed accurate expressions for the cooling
functions of H$_2$, HD and LiH in a wide range of density and
temperature that can be conveniently used in a variety of comological
applications.

\acknowledgements{It is a pleasure to thank T. Abel, A. Dalgarno, D.
Flower and P. Stancil for very useful conversations on various aspects
of the chemistry of the early universe. }

\appendix
\section{Appendix: Excitation Coefficients and Cooling}

We have evaluated the contribution of H$_2$, HD, and LiH to the 
heating/cooling properties of the primordial gas.
For each
molecule, the steady-state level populations are obtained at any $z$ by
solving the balance equations
\beq
x_{j'} \sum_{j'} (R_{jj'} + C_{jj'}) =\sum_{j'}x_{j'}(R_{j'j} +
C_{j'j}),
\enq
where $j$ and $j'$ indicate a generic couple of rovibrational levels.
The collisional transition probabilities $C_{jj'}$ and $C_{j'j}$ are
obtained by multiplying the corresponding excitation coefficients
$\gamma_{jj'}$ and $\gamma_{j'j}$ by the gas density. The terms
$R_{jj'}$ and $R_{j'j}$ are the radiative excitation and de-excitation
rates, that can be expressed in terms of the Einstein coefficients
$A_{jj'}$ and $B_{jj'}$,
\beq
R_{jj'}=\cases
{A_{jj'}+B_{jj'}u(\nu_{jj'},\tr), & $j'<j$,  \cr
B_{jj'}u(\nu_{jj'},\tr),          & $j'>j$,  \cr}
\enq
where 
$u(\nu_{jj'},\tr)$ is the energy density of the radiation per unit
frequency at the temperature $\tr$,
\beq
u(\nu_{jj'},\tr) = \frac{8\pi h \nu_{jj'}^3}{c^2}
\left[\exp(h\nu_{jj'}/k\tr)-1\right]^{-1}.
\enq

To evaluate the cooling function $\Lambda_{\rm mol}(z)$, the radiative
and collisional transition probabilities and the frequencies of the
roto-vibrational transitions must be specified for each molecule.
These data can be easily found in the literature only for H$_2$, that is
obviously the best studied case, but are quite sparse for the two
other species we have considered. As for LiH, the relevant quantities
have been collected and critically examined by Bougleux \& Galli (1997),
and we refer the interested reader to their paper (in particular,
see their Appendix B). Here, we limit ourselves to a discussion of 
the adopted coefficients for HD and H$_2$.

\subsection{H$_2$}
 
Ortho- and para-H$_2$ were assumed to be in their equilibrium ($3:1$)
ratio. The radiative rates were calculated by Turner et al. (1977).
Since in some cases where H$_2$ plays a major role the gas temperature
can reach more than one thousand degrees, it is not sufficient to
consider only the ground ($v=0$) vibrational state.  The choice of the
rovibrational H-H$_2$ collisional coefficients is very delicate, since
the published data may vary up to factors of 50--100 according to
different authors (see e.g.  Shull \& Beckwith 1982, Mandy \& Martin
1993, Sun \& Dalgarno 1994, Flower 1997a,b), and the resulting H-H$_2$
cooling functions may differ by considerable amounts. For cosmological
problems, one generally adopts the analytical expressions given by
Hollenbach \& McKee (1979, 1989) or Lepp \& Shull (1983).  These
formulae appear now to suffer considerably from the limitations and
uncertainties associated with the collisional coefficients available at
the time.  At low temperatures, the uncertainty is associated with the
choice of the interaction potential, since it is very difficult, even
nowadays, to calculate the potential in the interaction region to the
requisite level of accuracy ($\sim 10^{-3}$ hartrees). The recently
developed potentials LSTH (Truhlar \& Horowitz 1978), DMBE (Varandas et
al. 1987), BKMP1 (Boothroyd et al. 1991), BKMP2 (Boothroyd et al. 1996)
and PBSL (Partridge et al. 1993) have been adopted by various groups to
generate sets of collisional cross sections and rate coefficients,
using different methods.  Table~A1 shows a comparison of the resulting
values of $\gamma_{20}$ at $\tg=10^3$~K.

\begin{table}[t]
\caption{\sc H-H$_2$ Collisional Deexcitation: 
Comparison of Different Results at $\tg=10^3$~K}
\vspace{1em}
\begin{flushleft}
\begin{tabular}{llll}
\hline
Potential  &  Method  &  $\gamma_{20}$               & Reference \\
           &          &  (cm$^3$~s$^{-1}$)               &           \\
\hline
LSTH       &  rigid rotor     & $7.0\times 10^{-13}$ & GT  \\
LSTH       &  quasi-classical & $3.1\times 10^{-12}$ & MM  \\
LSTH       &  quasi-classical & $3.0\times 10^{-12}$ & LBD \\
DMBE       &  quasi-classical & $2.8\times 10^{-11}$ & LBD \\ 
DMBE       &  quantal         & $2.4\times 10^{-11}$ & SD  \\ 
BKMP1      &  quantal         & $1.3\times 10^{-12}$ & Fa  \\
PBSL       &  quantal         & $8.7\times 10^{-13}$ & Fb  \\
BKMP2      &  quantal         & $2.9\times 10^{-12}$ & FBDL \\
\hline
\end{tabular}
\vspace{1em}

\small{
GT: Green \& Thrular (1979);
MM: Mandy \& Martin (1993);
LBD: Lepp et al. (1995);
SD: Sun \& Dalgarno (1994);
Fa: Flower (1997a);
Fb: Flower (1997b);
FBDL: Forrey et al. (1997) 
}
\end{flushleft}
\end{table}

At temperatures higher than $\sim 600$~K, we have adopted the set of
rovibrational H-H$_2$ rate coefficients computed by Mandy \& Martin
(1993).  The corresponding cooling function has been determined by
Martin et al. (1996).  At lower temperatures, we have
adopted the collisional coefficients recently computed by Forrey et al.
(1997) that match those of Mandy \& Martin (1993) at $T=10^3$ K. We
have fitted the tabulated results by expressions like
\beq
\gamma_{J,J'}  =  a_0+a_1\tg+a_2\tg^2+a_3\tg^3,
\enq
where $\gamma_{J,J'}$ is in cm$^3$~s$^{-1}$, and the coefficients $a_i$
are listed in Table A2.  Different choices (like, e. g. Flower 1997a,b) 
are of course possible, and produce differences of a factor
$\sim 2$ in the cooling function.

\begin{table}[t]
\caption{\sc H-H$_2$ Collisional De-excitation Coefficients (from Forrey et al. 1997)}
\vspace{1em}
\begin{flushleft}
\begin{tabular}{llllll}
\hline
$J'$  &  $J$  &  $a_0$  &  $a_1$  &  $a_2$  &  $a_3$  \\
\hline
2     &  0    & 2.93(-14) & 1.21(-15) & 2.16(-19) & 1.32(-21) \\ 
3     &  1    & 8.34(-14) & 5.97(-16) & 7.76(-19) & 1.72(-21) \\
4     &  2    & 7.54(-14) & 2.65(-16) & 2.14(-19) & 2.65(-21) \\
5     &  3    & 2.95(-14) & 1.21(-16) & 5.53(-19) & 2.51(-21) \\
\hline
\end{tabular}
\end{flushleft}
\end{table}

It is convenient to express the H$_2$ cooling function $\Lambda_{{\rm
H}_2}$ (in erg~cm$^3$~s$^{-1}$) in the form given by Hollenbach \&
McKee~(1979),
\beq
\label{holl}
\Lambda_{{\rm H}_2}[n({\rm H}),\tg] = \frac{\Lambda({\rm LTE})}
{1+[n^{\rm cr}/n({\rm H})]},
\enq
where $\Lambda_{{\rm H}_2}({\rm LTE})$ is the LTE cooling function
given by Hollenbach \& McKee~(1979), $n^{\rm cr}$ is the critical
density defined as
\beq
\frac{n^{\rm cr}}{n({\rm H})}=
\frac{\Lambda_{{\rm H}_2}({\rm LTE})}{\Lambda_{{\rm H}_2}[n({\rm H})\rightarrow 0]},
\enq
and $\Lambda[n({\rm H})\rightarrow 0]$ is the low-density limit
of the cooling function, independent of the H density. The latter
can be computed from the collisional and radiative deexcitation
coefficients described above, and the result is well approximated 
by the expression
\begin{eqnarray}
\log \Lambda_{{\rm H}_2}[n({\rm H})\rightarrow 0] & = & 
-103.0+97.59\log\tg-48.05(\log\tg)^2    \nonumber \\
& + & 10.80(\log\tg)^3 -0.9032(\log\tg)^4,
\end{eqnarray}
over the range $10\;{\rm K}\leq \tg\leq 10^4\;{\rm K}$.

The H-H$_2$ cooling function is shown in Fig.~A1 for two values of the
H density, $n({\rm H})=0.1$~cm$^{-3}$ and $n({\rm H})=10^6$~cm$^{-3}$.
The cooling functions given by Hollenbach \& McKee (1989) and Lepp \&
Shull (1983), also shown in Fig.~A1, differ significantly from that
computed in this work especially in the low-density limit, at both low
and high temperatures (see discussion in Martin et al. 1996).

In considering the cooling by H$_2$ it is important not to neglect the
effect of reactive collisions of H$_2$ with H or H$^+$ that cause
ortho--para interchange and lead to a ratio out of the equilibrium
value 3:1. Calculations of the relative collisional coefficients have
been made for H$_2$--H$^+$ by Dalgarno et al.~(1973) and Flower \&
Watts~(1984), and for H$_2$--H by Sun \& Dalgarno~(1994). In the
conditions of the primordial gas with H$^+$/H $\ga 10^{-4}$ the
ortho--para exchange is dominated by collisions with protons, and
cooling at low temperatures may occur primarily through the
$J=2\rightarrow 0$ transition of para-H$_2$. A complete calculation of
the cooling function should take into account this effect that has been
neglected in the results presented in this section.

\begin{figure} 
\resizebox{\hsize}{!}{\includegraphics{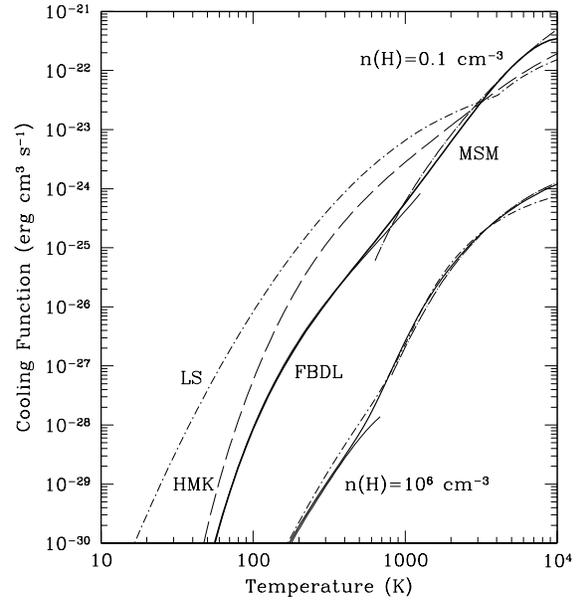}}
\caption[]{Cooling function of H$_2$ for H-H$_2$ collisions for two
values of the density. The thick solid curves show the results of our
fit as given by Eq. A5--A7.  The other curves give the cooling
functions computed by Lepp \& Shull (1983) (LS:  dot-dash), Hollenbach
\& McKee (1989) (HMK: dash), Martin et al. (1996) (MSM: dot-long dash),
and Forrey et al. (1997) (FBDL) (thin solid).}
\end{figure}

\subsection{HD}
 
Due to its small dipole moment, $D=8.3\times10^{-4}$ debyes (Abgrall et
al. 1982), the energy spacing of the rotational levels of HD is quite
large. The energy levels and the rotational radiative transition
probabilities have been computed by Dalgarno \& Wright (1972) from the
experimental data of Trefler and Gush (1968).  The calculations have
been extended by Abgrall et al. (1982), who considered both
dipole and quadrupole transitions and included a large number of
rotational and vibrational levels. The two sets of results differ by a
factor of two, and we have used the most recent determination.

The coefficients for inelastic scattering of He-HD have been computed
by Green (1974) using the Shafer \& Gordon (1973) potential of
H$_2$-He, and more recently by Schaefer (1990) with the Schaefer \&
K\"ohler~(1989) potential at temperatures $\tg \leq 600$ K for $0\leq J
\leq 3$ and $\Delta J=+1,+2$. De-excitation coefficients depend
approximately linearly with temperature.  We have fitted the numerical
data of Schaefer~(1990) with expressions of the form
$\gamma_{J'J}=a_0+a_1\;\tg^\alpha$, and fit coefficients are tabulated
in Table A3. These values should be divided by 1.27 to account for the
difference between the (HD, H) and (HD, He) interaction potentials
(Wright \& Morton 1979).

\begin{table}[t]
\caption{\sc HD-He Collisional De-excitation Coefficients (from Schaefer 1990)}
\vspace{1em}
\begin{flushleft}
\begin{tabular}{lllll}
\hline
$J'$  &  $J$  & $a_0$                &  $a_1$               &  $\alpha$    \\
\hline
1     &  0    & $4.4\times 10^{-12}$ & $3.6\times 10^{-13}$ &  0.77        \\
2     &  1    & $4.1\times 10^{-12}$ & $2.1\times 10^{-13}$ &  0.92        \\
2     &  0    & $3.4\times 10^{-13}$ & $1.1\times 10^{-14}$ &  1.10        \\
3     &  2    & $2.4\times 10^{-12}$ & $8.7\times 10^{-14}$ &  1.03        \\
3     &  1    & $3.2\times 10^{-13}$ & $1.3\times 10^{-15}$ &  1.47        \\
\hline
\end{tabular}
\end{flushleft}
\end{table}

\subsection{A total cooling function for the primordial gas}

In Fig. A2 we summarize the different contributions to the cooling of a
gas of primordial composition. Although the curves have been obtained
in the limits $n({\rm H})\rightarrow 0$, they are valid for $n({\rm
H})\la 10^2$~cm$^{-3}$.  At larger densities, the cooling functions
rapidly approach the corresponding LTE values; at intermediate
densities, the formula (\ref{holl}) represents a very good
approximation. The cooling functions for H-H$_2$ and H-HD are obtained
as described in this work, the H-LiH cooling function is computed
H$_2^+$ cooling function for collisions with H and e$^-$ was computed
by Suchkov \& Shchekinov (1978).

The cooling function of HD in the low-density limit for $\tg\la 10^3$~K
is accurately reproduced by the expression 
\begin{eqnarray}
\Lambda_{\rm HD}[n({\rm H})\rightarrow 0] & = &
    2\gamma_{10}E_{10}\exp(-E_{10}/k\tg) \nonumber \\
& + & (5/3)\gamma_{21}E_{21}\exp(-E_{21}/k\tg),
\end{eqnarray}
where $E_{10}=128 k$, $E_{21}=255 k$, and the $\gamma_{JJ^\prime}$ are
given in Table~A3.  As for LiH, the cooling function in the low-density
limit is well approximated by the polynomial expression
$\log\Lambda_{\rm LiH}(n{\rm H})\rightarrow
0)=c_0+c_1(\log\tg)+c_2(\log\tg)^2 +c_3(\log\tg)^3+c_4(\log\tg)^4$ in
the temperature range 10--10$^3$~K.  With $\Lambda_{\rm LiH}$ in erg
cm$^3$ s$^{-1}$, the fit coefficients are $c_0=-31.47$, $c_1=8.817$,
$c_2=-4.144$, $c_3=0.8292$, $c_4=-0.04996$.

In the standard model, the radiation temperature is higher than the
matter temperature and molecules are a net heating source for the gas.
Molecular hydrogen dominates the heating of the gas at temperatures
higher than $\sim 150$~K, whereas HD dominates at lower temperatures.
The ability of HD to heat/cool a gas when H$_2$ molecules become
unefficient is a remarkable property of a gas of primordial
composition, and has been verified in a number of studies (Varshalovich
\& Khersonskii 1977, Shchekinov 1986, Bougleux \& Galli 1997, Puy \&
Signore 1997).

\begin{figure} 
\resizebox{\hsize}{!}{\includegraphics{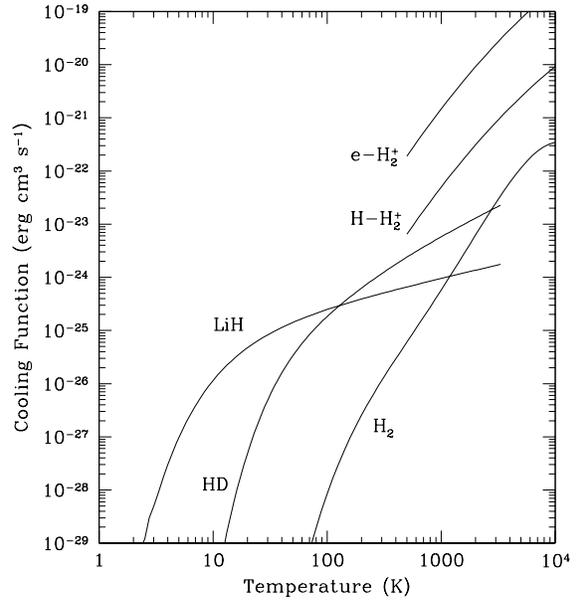}}
\caption[]{Cooling function per molecule of H$_2$, HD, LiH and H$_2^+$
in the low density limit ($n({\rm H})\la 10^2$~cm$^{-3}$).}
\end{figure}

\end{document}